\documentclass[pdflatex,sn-mathphys-num]{sn-jnl}

\usepackage{graphicx}%
\usepackage{multirow}%
\usepackage{amsmath,amssymb,amsfonts}%
\usepackage{amsthm}%
\usepackage{mathrsfs}%
\usepackage[title]{appendix}%
\usepackage{xcolor}%
\usepackage{textcomp}%
\usepackage{manyfoot}%
\usepackage{booktabs}%
\usepackage{adjustbox}
\usepackage{algorithm}%
\usepackage{algorithmicx}%
\usepackage{algpseudocode}%
\usepackage{listings}%
\usepackage{subcaption}
\usepackage{multirow}
\usepackage{array}
\usepackage{times}
\usepackage{pdflscape}
\usepackage{longtable}
\usepackage{tabularx}
\usepackage{placeins}  
\usepackage{siunitx}
\renewcommand{\arraystretch}{1.5}

\theoremstyle{thmstyleone}%
\theoremstyle{thmstyletwo}%

\theoremstyle{thmstylethree}%

\begin{document}

\title[Article Title]{Mental Effort Estimation in Motion Exploration and Concept Generation Design Tasks using Inter-Band Relative Power Difference of EEG}


\author*[1]{\fnm{G. Kalyan} \sur{Ramana}}\email{gkalyanramana@gmail.com}

\author[2]{\fnm{Sumit} \sur{Yempalle}}\email{sumityempalle@gmail.com}

\author[3]{\fnm{Prasad S.} \sur{Onkar}}\email{psonkar@des.iith.ac.in}

\affil[1, 2, 3]{\orgdiv{Department of Design}, \orgname{Indian Institute of Technology Hyderabad}, \orgaddress{\street{Kandi}, \city{Sangareddy}, \postcode{522084}, \state{Telangana}, \country{India}}}


\abstract{Conceptual design is a cognitively complex task, especially in the engineering design of products having relative motion between components. Designers prefer sketching as a medium for conceptual design and use gestures and annotations to represent such relative motion. Literature suggests that static representations of motion in sketches may not achieve the intended functionality when realised, because it primarily depends on the designers' mental capabilities for motion simulation. Thus, it is important to understand the cognitive phenomena when designers are exploring concepts of articulated products. The current work is an attempt to understand design neurocognition by categorising the tasks and measuring the mental effort involved in these tasks using EEG. The analysis is intended to validate design intervention tools to support the conceptual design involving motion exploration. A novel EEG-based metric, inter-Band Relative Power Difference (inter-BRPD), is introduced to quantify mental effort. A design experiment is conducted with 32 participants, where they have to perform one control task and 2 focus tasks corresponding to the motion exploration task (MET) and the concept generation task (CGT), respectively. EEG data is recorded during the 3 tasks, cleaned, processed and analysed using the MNE library in Python. It is observed from the results that inter-BRPD captures the essence of mental effort with half the number of conventionally used parameters. The reliability and efficacy of the inter-BRPD metric are also statistically validated against literature-based cognitive metrics. With these new insights, the study opens up possibilities for creating support for conceptual design and its evaluation.}

\keywords{Design neurocognition, Functionality, Motion exploration, EEG}



\maketitle

\section{Introduction}\label{sec1}



Product sketches have been well interpreted for their physical appearance and form \cite{prats2009transforming}, but communication of functionality still poses challenges \cite{wetzel2009automated, rizzuti2021interactive, dong2023towards}. This is especially true when the product contains articulated parts, i.e., parts having relative motion with respect to each other. The functionality of articulated products in such cases is a determinant of the kinematics and dynamics of the structure \cite{erdam1998mechanism}. However, static sketches of articulated product concepts do not give any dynamics-related information, such as its motion and forces acting on it,\cite{Babbage1826}, which is critical in understanding the functionality. For a designer to accurately depict an intended motion via a static sketch, it is taxing to the working memory \cite{SWELLER1988257}, which is parametrised as mental effort. Typically, mental effort to interpret/communicate the motion information gets translated into design actions such as annotations, gestures, etc. \cite{ramana2020designers} during the conceptual design phase. The effectiveness of this communication is directly proportional to the design outcome, especially in motion exploration \cite{onkar2013behaviour}. This study proposes a novel method to quantify designers' neurocognitive demands while sketching articulated product concepts. The method involves the empirical derivation of a novel metric to measure this demand in the form of the mental effort \cite{ramana2020designers} of the designer.

Cognitive processes such as working memory, attention, problem-solving, and decision making are intrinsically connected to the design process \cite{SWELLER1988257,gero2020framework}. Studies have also reported on the impact of the affective state of the designer on the design process \cite{SWELLER1988257, LIU2003341, dong2005latent, norman2007emotional, blessing2009drm}, which are often influenced by external or subjective factors that are difficult to control. Previous studies have established that designers translate their design intent (in the form of thoughts) into explicit design actions \cite{riva2011intention} during the conceptual design phase \cite{ramana2020designers,baule2016towards}, which implies that by tracking the designers' actions, one can objectively measure the design cognition \cite{hay2020future}. During this phase, a designer experiences a strained affective state and a higher demand on cognitive resources \cite{rizzuti2021interactive}. These factors may affect the decisions taken by a designer during the design process. Also, there is a long-term impact on the design of the product depending on the decisions taken during the conceptual design phase \cite{andreasen2015conceptual}.  Since cognitive factors are more objective and controllable and have direct implications for the outcomes of the design process, this study focuses on estimating the design cognition of the designers. This helps to identify hotspots in the design process that are high in cognitive demands for the designer. For instance, when a designer is sketching the concepts of a product, measuring the cognitive response for the whole sketching duration may help in identifying areas where the designer may experience a heightened mental effort. By designing appropriate tools or methods that specifically target reducing the mental effort of the designers during these hotspots, one may impact the efficacy of the design process.

Traditionally, protocol studies have been widely used to study designers' cognitive behaviour and related actions \cite{eastman1970analysis,Hay_Duffy_McTeague_Pidgeon_Vuletic_Grealy_2017,KAVAKLI_Gero2002}. However, protocol studies are subjective and prone to the limitations of the observer's interpretations \cite{li2021correlating,chiu_shu_2010,cross2001design}. Researchers have proposed alternative frameworks \cite{Gero_Milovanovic_2020} to measure design thinking \cite{brown2008design} through approaches of design physiology and design neurocognition. It is suggested that integrating the results obtained from these approaches can yield robust inferences on the design thinking and related processes  \cite{Gero_Milovanovic_2020}. In recent years, design neurocognition has emerged as a promising field to investigate design thinking and creative problem solving. It uses neurophysiological tools such as Electroencephalogram (EEG), functional Magnetic Resonance Imaging (fMRI), and functional Near-Infrared Spectroscopy (fNIRS) to study brain activity during real-world design tasks \cite{Hay_Duffy_McTeague_Pidgeon_Vuletic_Grealy_2017}. Due to technological advancements, research-grade EEG headsets are now available in the market as affordable wearable devices. This makes EEG technology easily accessible, portable and usable in the research community as compared to fMRI or fNIRS. 

The EEG technology has been used to measure, investigate or evaluate design from three broader perspectives - user \cite{zhu2023review}, process \cite{nguyen2015physiologically, Nguyen_2010,gero2020framework}, and outcome \cite{Li_Becattini_Cascini_2023, HuWanLin2017relationship}. In the context of \textit{research about design} \cite{frankel2010complex}, researchers are leveraging this technology to gather insights about the nature of the design process by evaluating the designers while they are designing, with the objective of facilitating the process much before the outcome is achieved. For the conceptual design phase, EEG has been related to various implicit phenomena of the design process such as creativity \cite{zangeneh2024eeg}, visual attention \cite{liang2017visual}, visual thinking \cite{YongZeng2010}, associative reasoning \cite{osti_10212428}, divergent thinking, convergent thinking and mental workload \cite{liu2018eeg}. However, there are limited studies that investigate the sketching activity of the designer for the conceptual design of articulated products, with a focus on motion exploration. In the context of neuroscience, EEG has been widely used to understand the nuances of motor imagery \cite{altaheri2023deep}, motion perception \cite{cochin1998perception}, movement parameters for motor control \cite{robinson2015adaptive}, which may be associated with the inherent ability of humans to mentally visualise motion. The need for studying motion exploration and mental effort for product concept generation has already been established in \cite{ramana2020designers}.

\subsection{Research gap}

This study investigates the mental effort involved while exploring the motion of articulated product concepts. In the concept generation task (CGT), designers visualise motion patterns of articulated product concepts using gestures, annotations, etc. in sketches. These motion patterns have to be realised by the appropriate embodiment of product components. For this, the designers have to analyse the motion associated with the structure through mental simulation. Such tasks are categorised as motion exploration tasks (METs) in this work. METs require a realistic, physics-based understanding and convergent thinking strategies, making them cognitively intensive. CGTs focus on ideating design concepts using methods like sketching or brainstorming, which require divergent thinking strategies \cite{blessing2009drm}. While CGTs are creativity-driven and may not always align with physical laws.METs and CGTs are complementary to each other because METs inform CGTs by grounding ideas in physical logic, while CGTs push the boundaries of ideation beyond immediate constraints. Current literature lacks objective, holistic measurements of mental effort during METs and CGTs. Protocol analyses, such as those in \cite{ramana2020designers}, are subject to the experimenter and labour-intensive, often resulting in low sample sizes that limit statistical validity. Subjective tools like NASATLX \cite{nikulin2019nasa} are useful but do not pinpoint critical cognitive hotspots, particularly during sketching. With advances in neuroscience, now is an opportune moment to explore EEG's potential in decoding neurocognition in design, a domain often seen as highly subjective and complex \cite{RAHARJO2008253,stolterman2008nature}.

EEG has proven valuable in mapping cognitive activity in design processes \cite{eastman1970analysis,cross2001design} and correlating neural patterns with design outcomes \cite{Li_Becattini_Cascini_2023}. However, as brain signals are multidimensional and complex to analyse a unified mathematical model relating brain signals and design cognition is essential to understand MET and CGTs.

\subsection{Research questions}
The current study aims to understand neurocognitive characteristic such as mental effort of the designer during a METs and CGTs. So, in this work, the following research questions (RQs) are formulated
\begin{itemize}
\item \textbf{RQ1}: \textit{Is there any difference in the mental effort involved in motion exploration and concept generation tasks? And if there is a difference, which one has the higher mental effort?}
\item \textbf{RQ2}: \textit{Is there any difference in the performance of the right and left brain lobes (defined by electrode locations) while performing motion exploration and concept generation tasks?}
\end{itemize}

\section{Methodology}
\label{sec:methodology}
\begin{figure}[!t]
    \centering
    \includegraphics[width=0.5\linewidth]{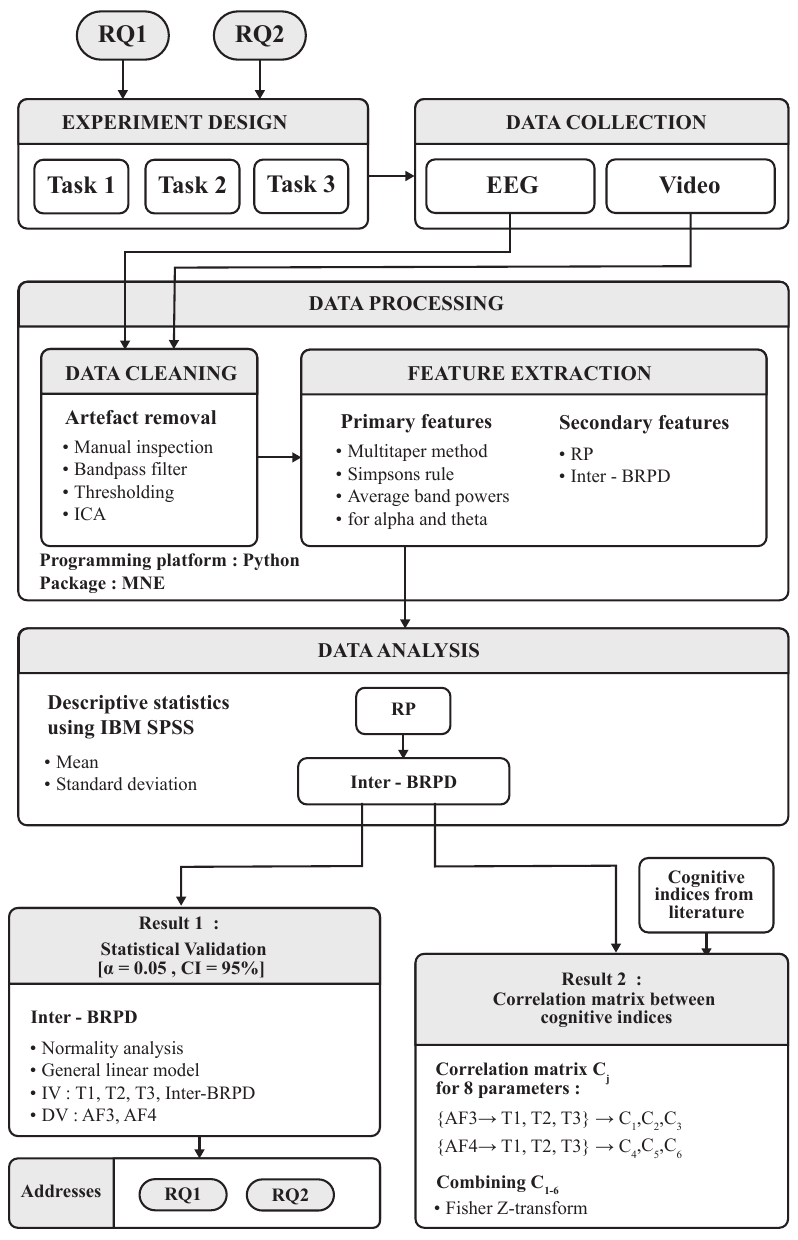}
    \caption{Methodology}
    \label{fig:methodology}
\end{figure}


From a measurement theory perspective \cite{measurement_theory_kjetil}, to validate the EEG-derived mental effort requires data capturing across multiple tasks. A single contrast between a control and one focus task may identify a change in mental effort, but it does not reveal variation across task types. Thus, including two cognitively demanding focus tasks, i.e. MET and CGT, alongside a control task allows for multi-point scaling, improving interpretability and robustness of the EEG measure. Furthermore, incorporating a low-mental effort control task provides a baseline essential for normalising EEG features, reducing inter-trial variability and enabling more stable intra-subject comparisons. From a statistical point of view, a three-condition within-subjects setup enables the use of repeated measures of ANOVA, which not only increases statistical power by controlling individual differences \cite{strale2024partitioning} but also supports multiple post-hoc contrasts. Thus, it preserves robustness against Type 1 errors through well-defined correction methods. 

\begin{figure}[!h]
    \centering
    \begin{subfigure}[t]{0.3\textwidth}
    \includegraphics[angle=-90,width=\textwidth]{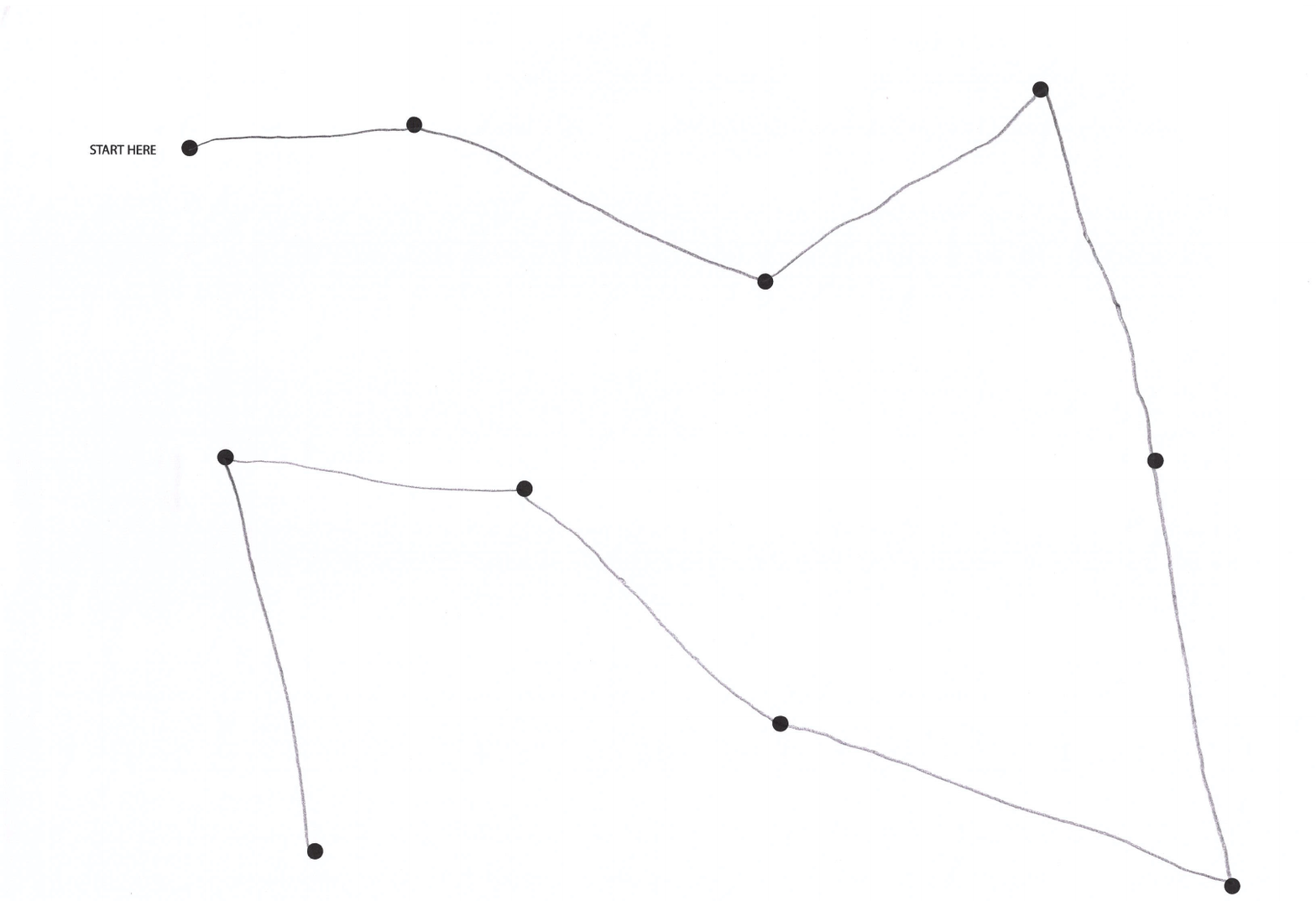}
        \caption{}
        \label{fig:T1}
    \end{subfigure}
    \hspace{0.5mm}
    \begin{subfigure}[t]{0.3\textwidth}
    \includegraphics[angle=-90,width=\textwidth]{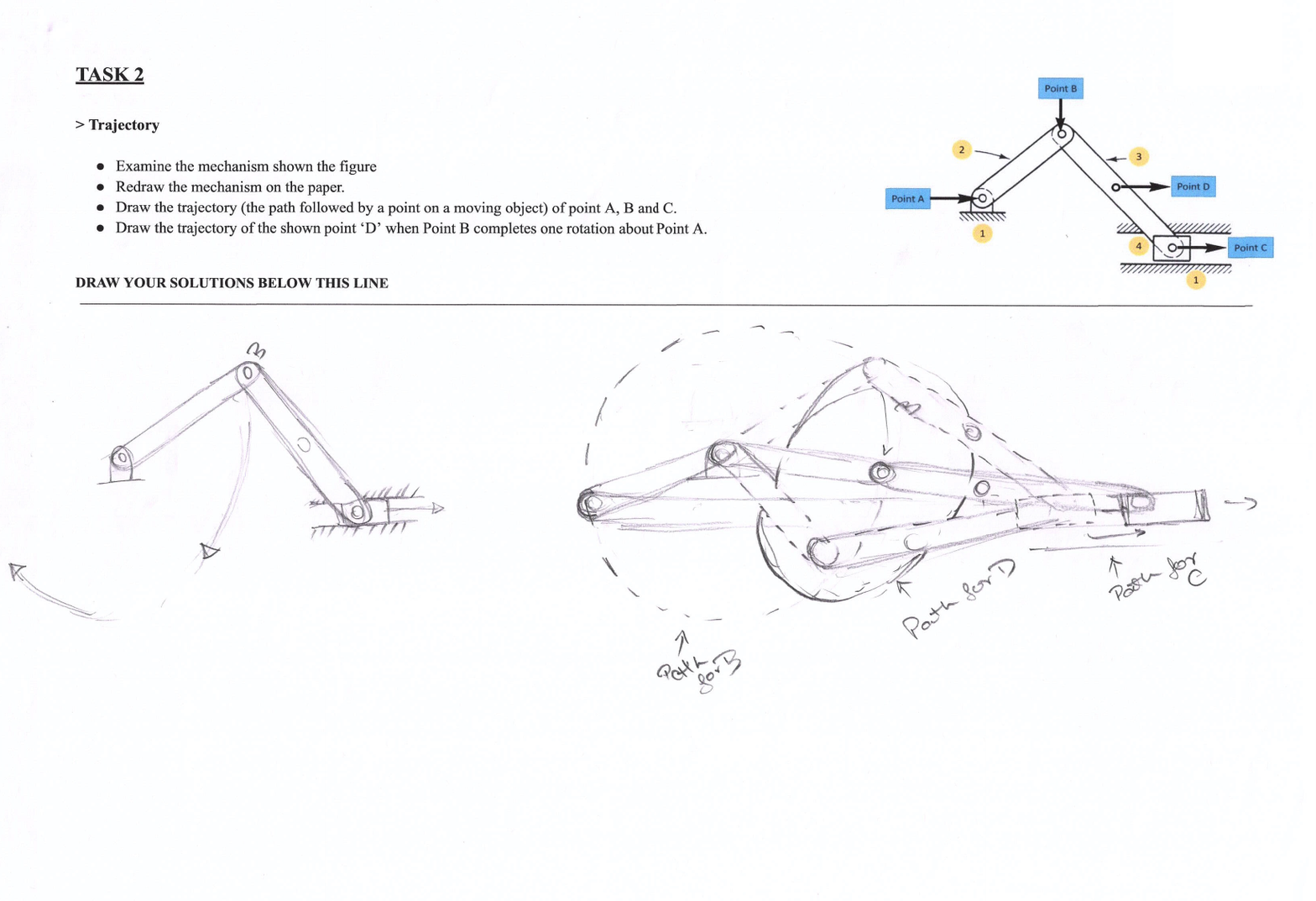}
        \caption{}
        \label{fig:T2}
    \end{subfigure}
    \hspace{0.5mm}
    \begin{subfigure}[t]{0.3\textwidth}
    \includegraphics[angle=-90,width=\textwidth]{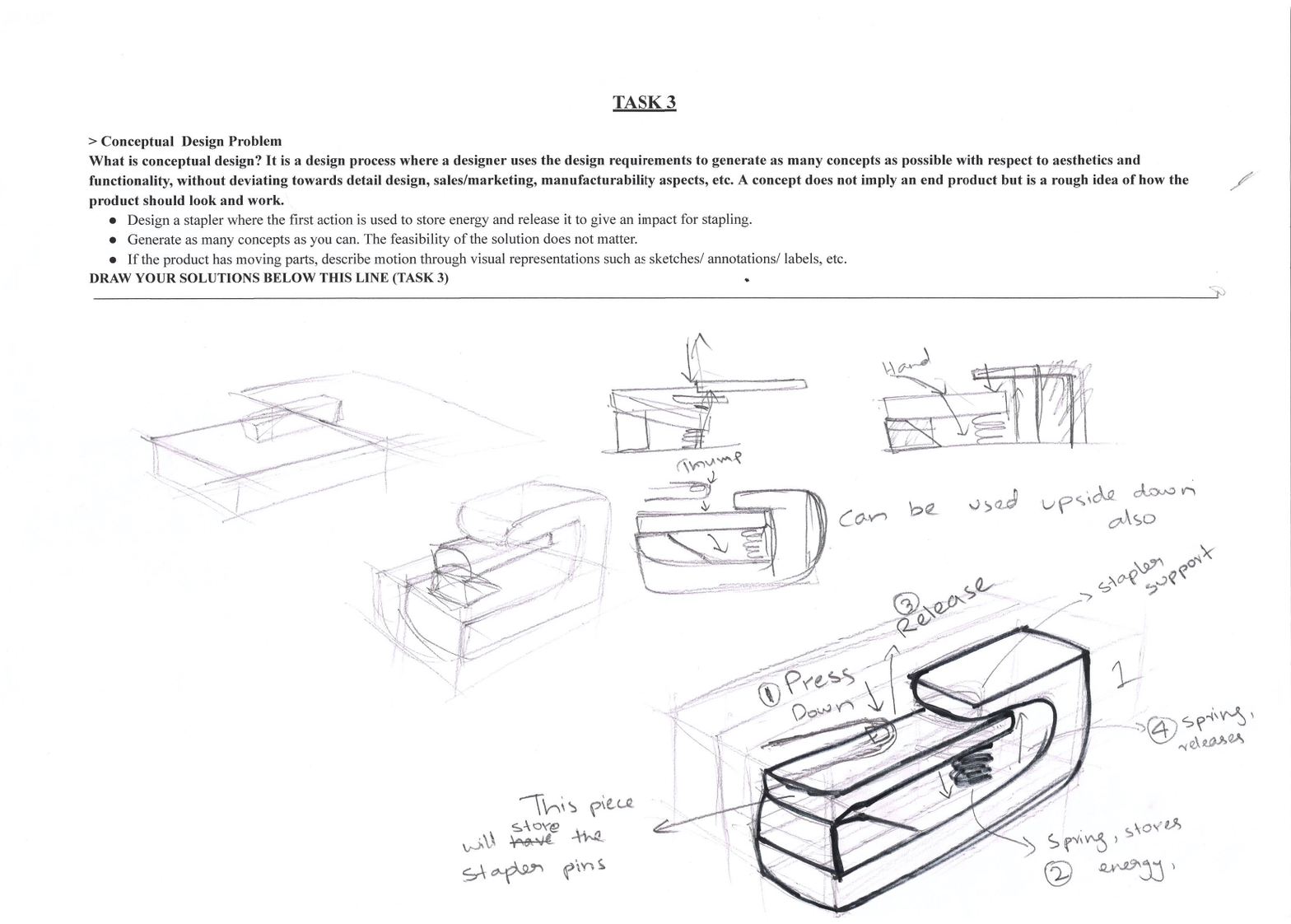}
        \caption{}
        \label{fig:T3}
    \end{subfigure}
    \caption{Sample responses of T1,T2 and T3. (a) The path traced by a participant connecting dots with a straight line. (b) Instances of the slider-crank mechanism to visualise the trajectory. (c) Articulated product concept sketch explaining its functionality.}
    \label{fig:Participant responses}
\end{figure}

Consequently, the methodological framework, shown in Fig. \ref{fig:methodology}, begins with a controlled experimental design comprising three sequential tasks (T1, T2, T3), aimed at addressing two core research questions (RQ1 and RQ2).  In this work, the complexity of a task is defined by the number of subtasks and the expected nature of their solutions. Task T1 has only one subtask, which does not require simultaneous mental processing of information and can have multiple solutions. Task T2 is an MET that has two subtasks, which include simultaneous spatial and temporal processing of information based on physics and has only one correct solution. Task T3 is a CGT, which includes subtasks of T2 with an added element of creative concept generation and can have multiple correct solutions. Hence, T2 is the most constrained with respect to the accuracy of the solution. The experiment is designed in such a way that the complexity increases from T1 to T3. The description for each task is given below : 
\noindent 
\begin{itemize}
   
    \item \textbf{T1 :} 
A spatial configuration of 10 randomly spaced black-coloured dots was given to the participants. Each participant received the same spatial configuration. One of the dots, referred to as the starting dot was labeled as ‘START HERE’. Participants were instructed to draw a straight line to connect the starting dot to the nearest next dot, without lifting the pencil until all dots were connected. To avoid repetition, it was instructed that two consecutive dots must be connected using a single stroke. Fig.\ref{fig:Participant responses}(a)	shows a sample response for T1.
 \item \textbf{T2 :} 
Fig. \ref{fig:Participant responses}(b) shows the answersheet for T2, which contains the task instructions, the question and the participant's response. Participants were requested to examine the mechanism shown in the question and redraw it. They were also asked to draw the trajectory of points A, B, C and D shown on the mechanism, when Point B completes one rotation about Point A. Participants sketched the response to the question in the sketching area as shown in the Fig. \ref{fig:Participant responses}(b) using a pencil or a marker.
 \item \textbf{T3 :}
Fig. \ref{fig:Participant responses}(c) shows the answersheet for T3 which contains the task instructions, the question and the participant's response. The problem statement given for T3 was as follows, \textit{‘Design a stapler where the first action is used to store energy and release it to give an impact for stapling’}. Participants were instructed to generate as many concepts as they could without worrying about the feasibility of the concepts. Participants sketched their responses in the sketching area using pencil and marker for T3 as shown in Fig. \ref{fig:Participant responses}(c) 
\end{itemize} 

The study assumes that as the complexity of the task increases, mental effort to perform the task also increases proportionately. It is also assumed that higher cognitive processing indicates higher mental effort. As the designer performs the tasks T1 to T3, his/her mental effort needs to be estimated for each task separately in order to answer RQs. The following hypothesis is formulated based on the RQs,

\vspace{\baselineskip}
\noindent \textbf{Hypothesis 1:}
\begin{itemize}
    \item \textbf{Null hypothesis ($H_0$)}: There is no difference between the tasks T1, T2 and T3 in terms of mental effort measured at electrode location AF3 and AF4. 
    \item \textbf{Alternate hypothesis ($H_a$)}: There is at least one difference between the tasks T1, T2 and T3 in terms of mental effort measured at electrode location AF3 and AF4. 
\end{itemize}

This essential characteristic of the mental effort needs to be recorded in the EEG data. Data were collected from multiple modalities: EEG recordings and video footage. EEG data were collected from only two electrode locations, AF3 and AF4, as these were the potential locations sensitive to mental effort. EEG data underwent comprehensive preprocessing to ensure signal integrity, involving manual artefact inspection, bandpass filtering, amplitude thresholding, and Independent Component Analysis (ICA) for artefact removal. All processing steps were implemented in Python using the MNE package, establishing a reproducible and scalable analysis environment. Following preprocessing, the signal was subjected to feature extraction. Primary features included multitaper spectral estimation, Simpson’s rule-based integration, and average band power calculation in the alpha and theta ranges. Secondary features, such as Relative Power (RP) and Band Relative Power Differences (BRPD), were derived to capture cognitive load dynamics.

To keep track of tasks, electrodes, and RPs is difficult, so new parameters have been introduced to reduce the dimensionality for easy understanding of the cognitive response of the participants. The inter-BRPD, especially its mean, is an indicative parameter of mental effort. 

Statistical analyses addressed two core results:
\begin{enumerate}
    \item Validation of BRPD metrics through general linear models and non-parametric tests (e.g., Wilcoxon signed-rank test)
    \item The construction of a cognitive correlation matrix across eight parameters, integrating EEG signals across tasks and channels (AF3, AF4) with Fisher Z-transformation.
\end{enumerate}
This methodological pipeline enables a robust investigation of the interplay between neurophysiological data and mental workload. Each major stage is explained in detail in the following sections. 

\subsection{Experimental Setup}
\subsubsection{Participants}
Participants for the experiment were recruited through an advertisement floated via email in the university student community. 32 university students  (M =24, F = 8) aged 20 to 45 years (M = 25.29, SD = 6.1) were selected for the experiment based on their educational background. Students from the design discipline or mechanical engineering who have completed at least two years of formal education in their respective domains were eligible for the experiment. Students with diagnosed atypical neurological conditions were excluded since they were beyond the scope of this study. Participants were selected using random sampling from the eligible pool of students. 29 participants were right-handed, and 3 participants were left-handed. 10 participants reported having prior music training. Participants were instructed not to consume any neurostimulants (tea, coffee) or alcohol 12 hours before the experiment. Each participant was given an e-gift voucher as incentives for their time and completion of the experiment.

\subsubsection{Tasks and apparatus}

\begin{figure}[htbp]
    \centering
    \begin{subfigure}[b]{0.49\linewidth}
        \centering
        \includegraphics[width=\linewidth]{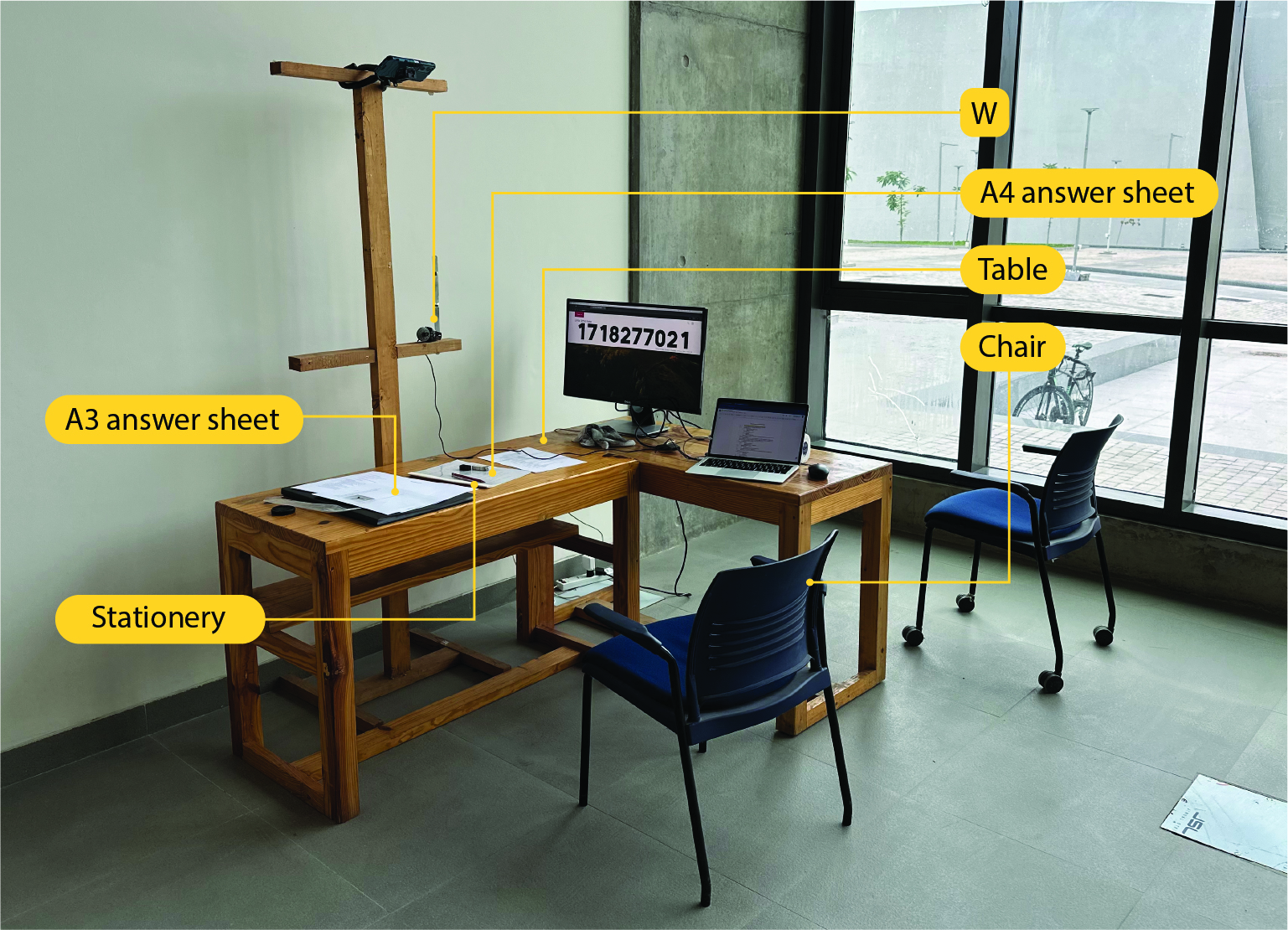}
        \caption{without participant}
        \label{fig: Experiment setup : (a) without participant}
    \end{subfigure}
    \hfill
    \begin{subfigure}[b]{0.49\linewidth}
        \centering
        \includegraphics[width=\linewidth]{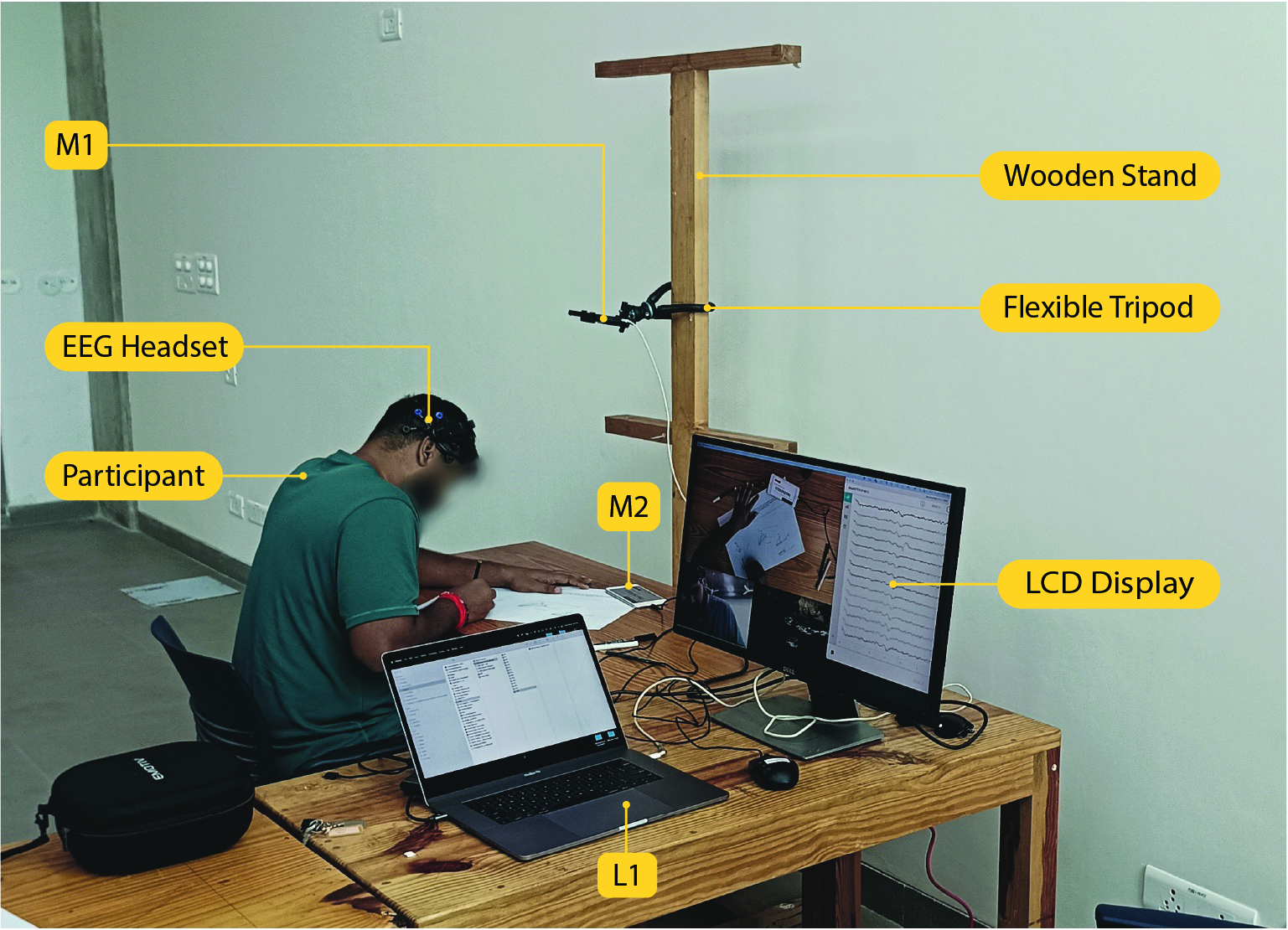}
        \caption{with participant}
        \label{fig: Experiment setup : (b) with participant}
    \end{subfigure}
    \caption{Experimental setup}
    
    \label{fig: Experimental setup : Experimental setup}
\end{figure}

All participants performed three tasks (T1,T2,T3), and each task included a sketching activity. Before starting the experiment, participants were asked to perform a practice task that introduced them to the theme of motion-exploration through simple questions related to motion visualisation. This ensured that the participants were not subjected to a surprise element during the task, which could impact the EEG recordings. No EEG data were recorded for this task; responses to this task were not included in data analysis. T1, T2 and T3 were performed on an A3-size paper, respectively. 

Participants were invited individually to perform the sketching tasks inside an isolated room. Experimenters explained the procedure to each one, after which participants provided consent for data use. The participants were allowed to opt out of the experiment at any point of time. The experimental setup is shown in Fig. \ref{fig: Experimental setup : Experimental setup}(a), which shows a chair and a table provided to the participant to sit and perform the sketching tasks. Stationery and a set of printed answer papers, which included task instructions and fixation marks, were kept in a predetermined order in front of the participant on the table. A webcam (W) was also placed to record the video of the participant's face. Fig. \ref{fig: Experimental setup : Experimental setup}(b) shows devices used for data collection – two mobile phones (M1 and M2), a Macbook Pro laptop (L1), an LCD Display and a 14-channel Emotiv® EPOC X wireless EEG headset for EEG data collection, a flexible tripod and a wooden stand. Additionally, Biotrue lens cleaning solution and tissue paper roll were also used in the data collection process.

\subsection{Data collection}
\label{Data collection}

The following data were collected throughout the experiment:  
\begin{itemize}
    \item EEG data for T1, T2 and T3
    \item Sketching video for T1, T2 and T3
    \item Video of the faces of participants for all three tasks
    \item Scans of sketches of the participants for T1, T2 and T3
    \item Experiment feedback form
\end{itemize}

M1 was used to record the video of the sketching activity from a top-view, and M2 was used to continuously display Unix \cite{Moreno2020} time. In order to synchronise the EEG recording with the sketching activity, it was ensured that the Unix time displayed on M2 was always captured in the video recording by M1. Webcam W was placed on the table to capture a video of the participant's face. 

EEG non-invasively captures brain activity via scalp electrodes \cite{teplan2002fundamentals}, measuring time-varying voltage signals from neuronal activation \cite{faber2018two, BALL2009708}. These signals are analysed across delta (1–4 Hz), theta (4–8 Hz), alpha (8–13 Hz), and beta (13–30 Hz) bands \cite{yuvaraj2014analysis}. The theta \cite{Wascher2013} and alpha \cite{Bacigalupo2022} bands are especially sensitive to cognitive states like attention, fatigue, and engagement \cite{blanco2024real}.

To interpret the brain activity, raw EEG signals need to be collected, cleaned, processed and analysed. EEG data were recorded using the 14-channel Emotiv® Epoch X device at a sampling rate of 256 Hz and the international 10-20 system for electrode placement. Referential montage with a common mode sensor (CMS) as the left mastoid and driven right leg (DRL) as the right mastoid was used to enable a cleaner EEG signal with common mode rejection. The electrode locations AF3 and AF4 \cite{mahajan2017real, astuti2024investigating} in the frontal lobe were chosen for examining the effects of the tasks on the brain activity of the designer. The frontal-alpha and frontal-theta will be referred to as alpha and theta for the rest of the paper. Data was stored in EDF format on the local hard disk of a computer. 

Fig. \ref{fig: Data collection: Data collection for Task n (where n=1,2,3)} shows the process followed for EEG data collection. Data was recorded separately for all three sketching tasks. Each task recording included a baseline EEG and a task-related EEG, recorded continuously without any pause in between. Baseline EEG was recorded for the first 30 seconds of every task. For the first 15 seconds of the baseline period, the participant was asked to look at a fixation ‘x’ mark printed on the centre of an A3 sheet and relax with their eyes open. For the next 15 seconds, the participant was asked to close their eyes and relax. After this, the fixation mark was removed, and the participant was instructed to begin the sketching task immediately. A maximum time limit was mentioned for each task in the task instructions. However, the participants were allowed to take as much time as they wished to complete the tasks. The EEG recording was stopped after the participant indicated their task completion. This process was repeated for all the subsequent tasks. After completing task 3, the EEG headset was removed from the head of the participant, and he/she was asked to fill out the experiment feedback forms in Google Forms.

\begin{figure}[H]
    \centering
    \includegraphics[width=0.4\linewidth]{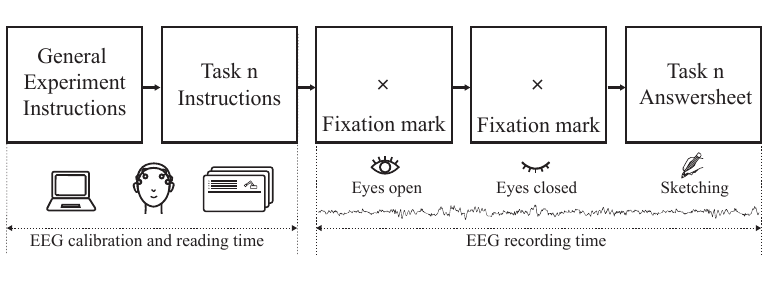}
    \caption{Data collection for Task n (where n=1,2,3)  }
    \label{fig: Data collection: Data collection for Task n (where n=1,2,3)}
\end{figure}

\subsection{Data Processing}
The amplitude of the scalp EEG voltage signal typically fluctuates within a few microvolts with a very high temporal resolution. This makes EEG signals susceptible to various kinds of noise and artefacts, which need to be eliminated while minimally affecting the signal of interest. Artefacts are parts of the recorded signal that arise from sources other than neural activity in the brain. As mentioned in \cite{islam2016methods}, the sources of the artefacts could be in the environment around the subject, the measuring instrument itself and/or the biological processes of the subject under observation. In certain experimental setups, endogenous neural signals may be classified as artefacts, depending on the research objectives. For instance, for a study investigating the brain's response to an auditory stimulus, neural activity recorded during pressing a button in response to the auditory stimulus may be considered an artefact.  Despite originating from legitimate cerebral processes, these motor-related neural activities could potentially hide the target sensory response. This emphasises the importance of clearly defining the signal of interest in neurophysiological research. It implies that the classification of neural activity as either a relevant signal or an artefact is not absolute, but rather depends on the specific research question being addressed. 

The current study aims to examine the cognitive response of the designers' brains while performing tasks with varying levels of complexity. For the sake of simplicity, this study assumes that the tasks presented to the designer are the stimuli for which the gross response of the brain in the form of mental effort is of interest to the authors. It does not focus on evaluating the subject's neural responses to any specific sensory stimulus, such as an auditory or haptic stimulus that may be presented during the tasks. 
\subsubsection{EEG Data cleaning}
The key artefacts, as mentioned in \cite{islam2016methods}, addressed in this study are electrocardiogenic (ECG), electrooculogenic (EOG) and electromyogenic (EMG). Researchers have implemented several artefact removal techniques \cite{islam2016methods} based on independent component analysis (ICA), wavelet transform, deep learning, and manual rejection to clean the EEG signals. Several tools like EEGLAB, FieldTrip, Brainstorm, BCILAB based on MATLAB; MNE, NeuroKit2, BioSPPy, PyEEG based on Python and OpenViBE based on  C++ are used to clean artefacts and process EEG signals \cite{unakafova2019comparing}. This study uses a combination of ICA and the manual method to eliminate the artefacts. The EEG data is processed using the Python programming language \cite{vanrossum2009python} and the MNE library \cite{gramfort2013meg}. Post this, features are extracted from the cleaned signals to derive meaningful interpretations.

\begin{figure}[H]
    \centering
    \includegraphics[width=0.4\linewidth]{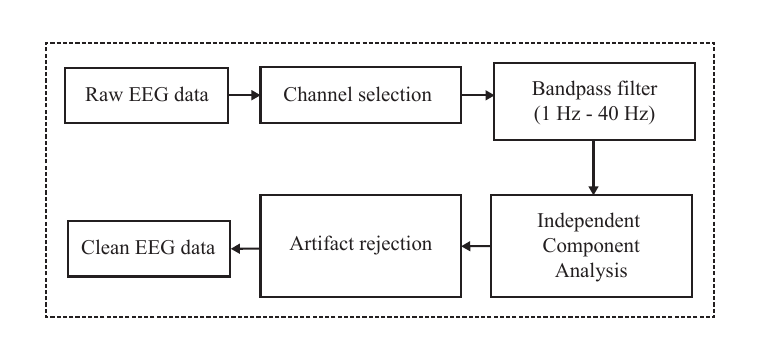}
    \caption{EEG data cleaning}
    \label{fig: EEG Pre-processing: EEG data cleaning}
\end{figure}

From each participant EEG recordings were collected for 3 tasks. Out of 32 data sets, 5 datasets were removed due to faulty EEG channels or failure of the EEG headset during the experiment. Effectively, 27 data sets were taken for cleaning. In order to remove the artefacts and clean the EEG signals, the authors followed the pipeline mentioned in Fig. \ref{fig: EEG Pre-processing: EEG data cleaning} using the MNE Python library.   

After selecting the channels, raw EEG data is filtered using a digital FIR bandpass filter with  1 and 40 Hz as cut-off frequencies. The filter parameters are shown in Table \ref{tab:filter_specifications}. Since Emotiv EPOC X is a wireless device that transmits EEG data using Bluetooth, a notch filter corresponding to the line noise of 50 Hz was not needed.

\begin{table}[htbp]
\centering
\caption{Filter Specifications for High-Pass and Low-Pass Filters}
\label{tab:filter_specifications}
\begin{tabular}{@{}lcc@{}}
\toprule
\textbf{Parameter} & \textbf{High-Pass Filter} & \textbf{Low-Pass Filter} \\
\midrule
Window Type & Hamming & Hamming \\
Passband Ripple & 0.0194 & 0.0194 \\
Stop-Band Attenuation & 53 dB & 53 dB \\
Passband Edge & 1.00 Hz (lower) & 40.00 Hz (upper) \\
Transition Bandwidth & 1.00 Hz (lower) & 10.00 Hz (upper) \\
\(-6\,\text{dB}\) Cut-Off Frequency & 0.50 Hz & 45.00 Hz \\
Filter Length & 845 samples (3.301 s) & 85 samples (0.332 s) \\
\bottomrule
\end{tabular}
\end{table}
 
After filtering, the EEG signal is inspected manually for very large amplitude signals (above ±200 µV) corresponding to body movement artefacts. The facial video, which was captured using the webcam (W), is used to correlate very large amplitude EEG signals with the head and neck movements. These signal portions are annotated as `bad signals'. The bad signals are rejected implicitly during the ICA. For each EEG recording, out of the 14 channels, only the frontal 8 channels (AF3, F7, F3, FC5, FC6, F4, F8, AF4) were selected for implementing the ICA \cite{zhou2019many} for optimal classification performance of the independent components. ICA was performed using the FASTICA algorithm \cite{gramfort2014mne} (with a maximum iteration limit kept at 15000) to identify the ECG, EOG and EMG artefacts. ECG and EOG artefacts are rejected from the signal to obtain a clean EEG signal using the methods mentioned in \cite{uriguen2015eeg}. EMG artefacts were identified using the muscle-artefact identification algorithm of the MNE library and manual inspection.  

\subsubsection{Feature extraction}
\label{Feature extraction}
The EEG signals can be characterised using time domain features \cite{Alfahoum2014}, frequency domain features \cite{chen2023eeg} or a combination of both \cite{MORALES2022101067}. The high temporal resolution of EEG signals makes it difficult to rely solely on the time domain features to identify any conclusive patterns. However, for longer duration of studies that last for several minutes, the frequency spectrum of EEG signals has rich data, which may be useful for predicting some patterns. EEG signals are commonly analysed across four distinct frequency ranges, known as brain waves: delta (1–4 Hz), theta (4–8 Hz), alpha (8–13 Hz), and beta (13–30 Hz) \cite{yuvaraj2014analysis}. The theta \cite{Wascher2013} and alpha \cite{Bacigalupo2022} bands have been suggested to be sensitive to cognitive factors such as mental fatigue, attention and engagement \cite{blanco2024real}. The spectral power computed over the range of frequencies within a band, called as average band power (ABP) or band power (BP), is a common metric used to derive meaningful information about the brain activity for a specific task \cite{klimesch1999eeg,kim2022exploring,nguyen2015physiologically,Stancin2021,YongZeng2010}. This implies that, for a given task, the variation in the power of a given frequency band observed at an electrode location is indicative of the cognitive response of the brain. 

Event-related power fluctuations such as event-related synchronisation (ERS) and event-related desynchronization (ERD) have been widely referred for studying the brain activity for stimulus-based tasks from a microscopic perspective \cite{klimesch1999eeg}. Some studies have also considered a macroscopic view of time to observe power fluctuations throughout the duration of the task \cite{gerner2025neurophysiological, hu2022design}. A suppression in the power of the alpha band, also called alpha desynchronization, is a known marker of cognition and heightened attention \cite{klimesch1999eeg}. An increase in the power of frontal theta is observed with an increase in working memory load \cite{jensen2002frontal} and task complexity \cite{sammer2007relationship}.

This study follows the macroscopic approach of looking at time with an objective of studying the gross implications of the power changes towards the cognitive response in terms of mental effort of the designer. It primarily investigates the changes in theta and alpha band power for a given task.

\begin{figure}[t]
    \centering
    \includegraphics[width=0.6\linewidth]{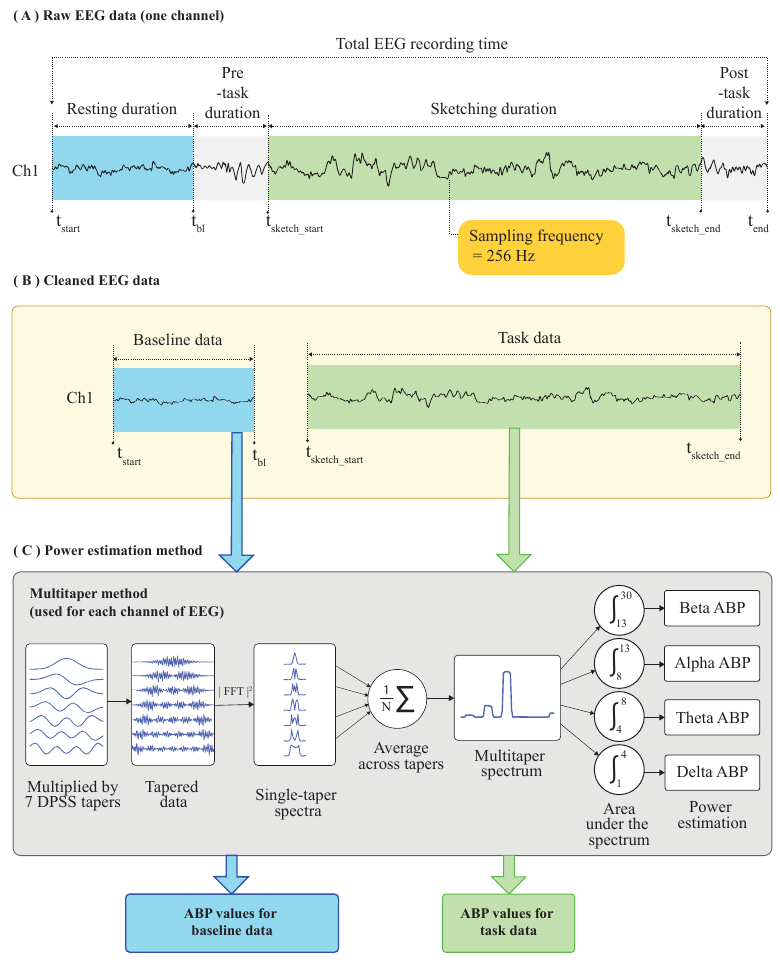}
    \caption{EEG feature extraction}
    \label{fig: EEG signal timing diagram}
\end{figure}

In Fig. \ref{fig: EEG signal timing diagram}, (A) shows the timing diagram for an example channel Ch 1. The EEG recording begins at time $t_{start}$. The duration between $t_{start}$ and $t_{bl}$ is the resting duration where the participant is not performing any sketching task. Hence, the EEG features computed for this time interval are considered as a reference for measuring any changes induced by the tasks. The duration between $t_{sketch\_start}$ and $t_{sketch\_end}$ is the sketching duration where the participant is performing the sketching tasks, hence, EEG features computed for this interval are assumed to be indicative of effects of the task on the cognitive response of the participant.

In order to extract the features, the cleaned EEG signal data is first cropped into two segments called the baseline data and the task data, as shown in (B) of Fig. \ref{fig: EEG signal timing diagram}. Features are extracted for both segments using the same methods as described further. Firstly, the Power Spectral Density (PSD) for both segments was estimated using the multitaper method for frequency bins up to 128 Hz as per the Nyquist criteria. In the multitaper technique, the original signal is first passed through a set of optimal bandpass filters called Discrete Prolate Spheroidal Sequences (DPSS) or Slepian sequences \cite{Slepian1978}. This filtering is achieved by convolving the Slepian sequences with the original signal. After filtering, a periodogram is computed for each of these new filtered signals. The final spectrum is derived by averaging across all the periodograms. As the Slepian sequences are orthogonal to one another, this approach provides statistically independent estimations of the underlying spectrum. The multitaper method has been proven to reduce the spectral leakage and variance in PSD estimation as compared to the Welch method or the FFT method, which are traditionally used for PSD estimation \cite{mitra1999analysis, Babadi2014, demirel2021single}. Secondly, the area under the PSD curve was calculated using Simpson's rule of integration to estimate the average band power for four frequency bands - delta, theta, alpha and beta. Two sets of ABP values for each frequency band were determined as primary features, one for the baseline data and other for task data. An example of such primary features are shown in Table \ref{tab:parameters_estimation} from step 1-5. Secondary features, such as brain ABP, RPs and inter-BRPDs, are derived using the average band powers. RPs are calculated as shown in Table \ref{tab:parameters_estimation} in steps 7 and 8. Inter-BRPD calculation is shown in step 9. 

\begin{table}[t]
\centering
\scriptsize
\caption{Data processing to obtain relative band powers}
\label{tab:parameters_estimation}
\renewcommand{\arraystretch}{1.4}
\begin{tabular}{p{1cm} p{7cm} p{2cm}}
\toprule
\textbf{Step No.} &\textbf{Measure} & \textbf{Value} \\
\midrule
1&Alpha ABP ($\alpha_{ABP}$) & 1.72E-11 \\
2&Alpha baseline ($\alpha_{b}$) & 3.36E-11 \\
3&Theta ($\theta_{ABP}$) & 2.46E-11 \\
4&Theta baseline ($\theta_{b}$) & 1.72E-11\\
5&Baseline Brain ABP (B$_b$) & 1.29E-10 \\
6&Brain ABP (B$_{ABP}$) =$\delta_{ABP}$+$\theta_{ABP}$+$\alpha_{ABP}$+$\beta_{ABP}$ & 1.43E-10 \\
7&Alpha Relative Power (\%)= $\alpha_{rel}$ = $\alpha$ / B$_{ABP}$ × 100 & 14.69 \\
8&Theta Relative Power (\%) = $\theta_{rel}$ = $\theta$ / B$_{ABP}$ × 100 & 21.18 \\
9&Inter-Relative Band Power Difference (inter-BRPD) = $\alpha_{rel}$ $-$ $\theta_{rel}$ & -6.48 \\
\bottomrule
\end{tabular}
\end{table}

\subsection{Data Analysis}
The band relative powers give the contribution of the same in the ABP for the sketching duration. To analyse the relative power data, in the present study, the bands alpha and theta are considered, as they are sensitive to cognitive load. Bands delta and beta are discarded for the study. Relative powers are estimated at two electrode locations (AF3 and AF4) and for three tasks, T1-T3. This implies that for every task and every electrode, RPs for alpha and theta are obtained. There are two types of analyses followed in the present study, i.e. taskwise and electrode-wise, to get a meaningful understanding of the participants' response in terms of RPs of alpha and theta to the tasks T1-T3. 

\subsubsection{Relative power analysis}
Taskwise analysis implies keeping the electrode location and band the same; the mean RP is compared between tasks T1, T2 and T3. Statistically speaking, this is termed as ``between-subjects". Electrode-wise analysis implies keeping the task and band the same, RP is compared between the electrode locations. Electrode-wise analysis gives a meaning of whether there is a difference between the data obtained from electrode locations AF3 and AF4. Statistically speaking, this is termed as ''within-subjects".  

\begin{figure}[t!]
    \centering
    \begin{subfigure}[t]{\linewidth}
        \centering
       \includegraphics[height=7.5cm]{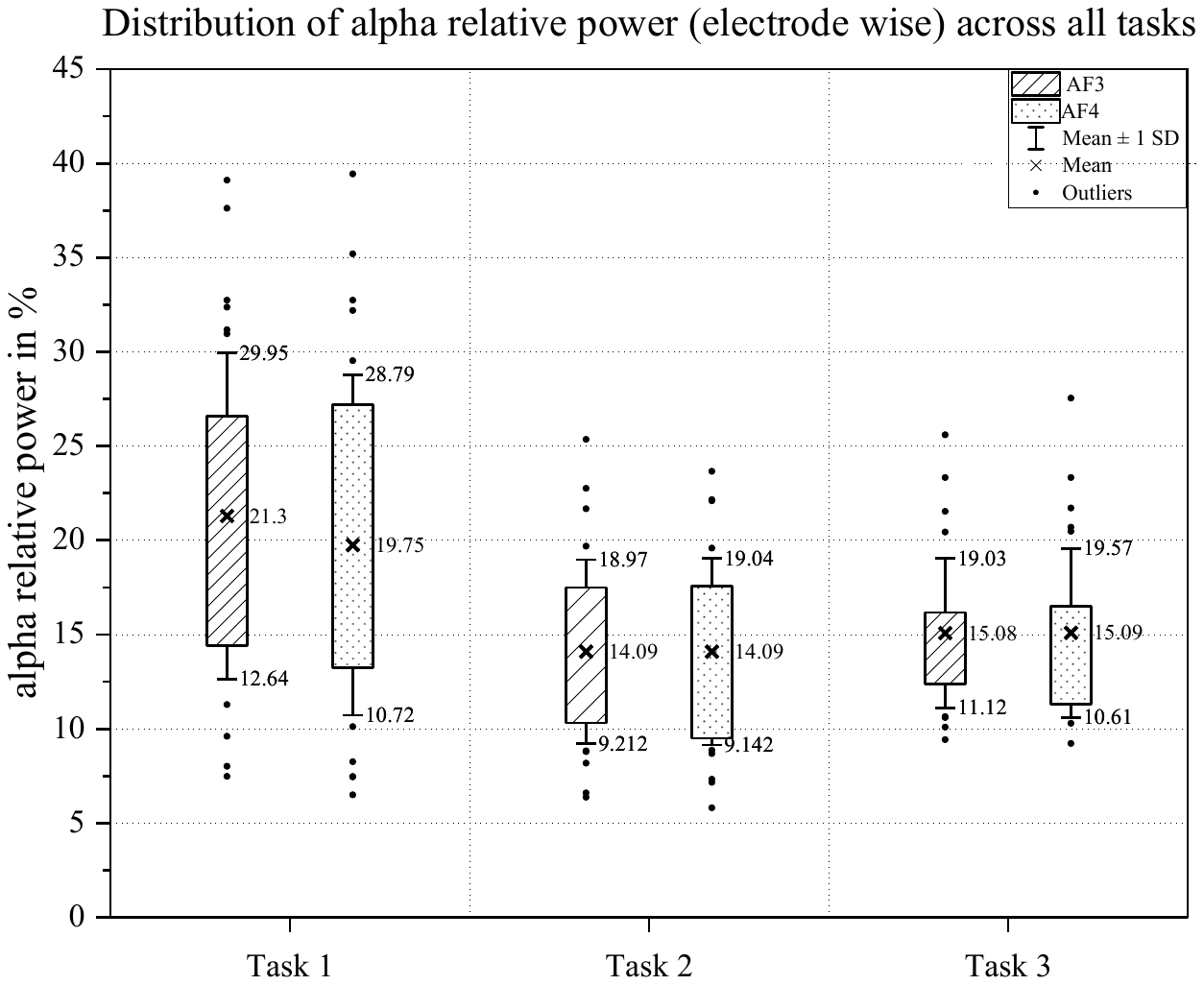}
        \caption{}
    \end{subfigure}%
    
   \begin{subfigure}[t]{\linewidth}
        \centering
        \includegraphics[height=7.5cm]{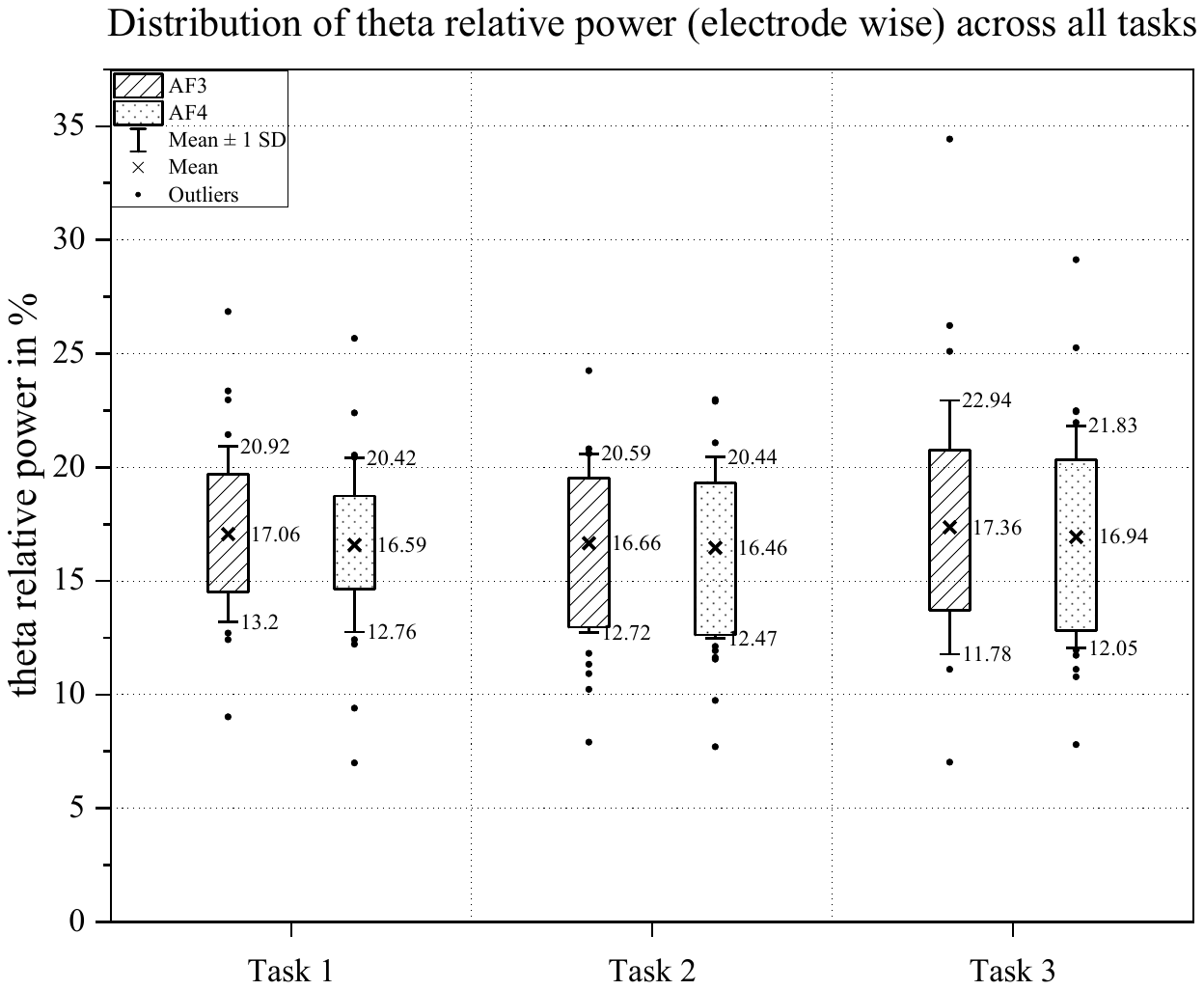}
        \caption{}
    \end{subfigure}
    \caption{Distribution of RP (electrode-wise) across all tasks (a) alpha band (b) theta band}   
    \label{fig:graph_distribution_alpha_theta}
\end{figure}

\begin{table}[htbp]
\small
\centering
\caption{Descriptive Statistics of Relative Power by Electrode, Task, and Band}
\footnotesize
\begin{tabular}{lll
S[table-format=2.2]
S[table-format=2.2]
S[table-format=2.2]
S[table-format=2.2]
S[table-format=2.2]
S[table-format=2.2]
S[table-format=2.2]
S[table-format=2.2]}
\toprule
\textbf{Electrode} & \textbf{Task} & \textbf{Band} &
\textbf{Mean} & \textbf{CI Low} & \textbf{CI High} & \textbf{Median} &
\textbf{SE} & \textbf{SD} & \textbf{Min} & \textbf{Max} \\
\midrule
\multirow{6}{*}{AF3} 
  & \multirow{2}{*}{T1} & Alpha & 21.30 & 17.87 & 24.72 & 20.30 & 1.66 & 8.65 & 7.48 & 39.11 \\
  &                    & Theta & 17.06 & 15.53 & 18.59 & 16.86 & 0.74 & 3.86 & 9.02 & 26.85 \\
  & \multirow{2}{*}{T2} & Alpha & 14.09 & 12.16 & 16.02 & 13.77 & 0.94 & 4.88 & 6.37 & 25.35 \\
  &                    & Theta & 16.66 & 15.10 & 18.21 & 17.66 & 0.76 & 3.93 & 7.91 & 24.25 \\
  & \multirow{2}{*}{T3} & Alpha & 15.08 & 13.51 & 16.64 & 14.70 & 0.76 & 3.95 & 9.42 & 25.60 \\
  &                    & Theta & 17.36 & 15.15 & 19.57 & 16.39 & 1.07 & 5.58 & 7.02 & 34.43 \\
\midrule
\multirow{6}{*}{AF4} 
  & \multirow{2}{*}{T1} & Alpha & 19.75 & 16.18 & 23.32 & 18.36 & 1.74 & 9.03 & 6.49 & 39.43 \\
  &                    & Theta & 16.59 & 15.07 & 18.10 & 17.31 & 0.74 & 3.83 & 6.99 & 25.67 \\
  & \multirow{2}{*}{T2} & Alpha & 14.09 & 12.13 & 16.05 & 15.01 & 0.95 & 4.95 & 5.81 & 23.66 \\
  &                    & Theta & 16.46 & 14.88 & 18.03 & 17.43 & 0.77 & 3.99 & 7.70 & 22.97 \\
  & \multirow{2}{*}{T3} & Alpha & 15.09 & 13.32 & 16.86 & 14.45 & 0.86 & 4.48 & 9.22 & 27.54 \\
  &                    & Theta & 16.94 & 15.00 & 18.87 & 16.42 & 0.94 & 4.89 & 7.80 & 29.13 \\
\bottomrule
\end{tabular}
\label{tab:desc_stats_RP}
\end{table}

\begin{table}[h!]
\centering
\caption{Difference (AF3 $-$ AF4) of mean RPs}
\label{tab:AF3-AF4}
\scriptsize
\begin{tabular}{llccc}
\toprule
\textbf{Band} & \textbf{Task} & \textbf{AF3} & \textbf{AF4} & \textbf{AF3$-$AF4} \\
\midrule
\multirow{3}{*}{Alpha} 
 & T1 & 21.29 & 19.75 & 1.54 \\
 & T2 & 14.09 & 14.09 & 0.00 \\
 & T3 & 15.08 & 15.09 & -0.01 \\
\midrule
\multirow{3}{*}{Theta} 
 & T1 & 17.06 & 16.59 & 0.47 \\
 & T2 & 16.66 & 16.46 & 0.20 \\
 & T3 & 17.36 & 16.94 & 0.42 \\
\bottomrule
\end{tabular}
\end{table}

\begin{enumerate}
\item \textbf{Taskwise analysis}: The graph contains box plots of the RP of alpha (Fig. \ref{fig:graph_distribution_alpha_theta}(a)) and theta (Fig. \ref{fig:graph_distribution_alpha_theta}(b)) for each task for each electrode. The descriptive statistics are shown in Table \ref{tab:desc_stats_RP}.

\begin{itemize}
    \item \textbf{Alpha band}: Fig. \ref{fig:graph_distribution_alpha_theta}(a) shows alpha RP across all electrodes and tasks. Considering electrode location AF3 and band alpha, task T2 has the lowest value, and T1 has the highest. For task T3, the mean alpha RP   is between that of T1 and T2. This shows that for electrode location AF3 and the band alpha, there is a U-shaped relation between the RP across tasks T1, T2 and T3.

    Consider the electrode location AF4 and the alpha band; T2 has the lowest value, and T1  has the highest. For T3, the mean alpha RP is between that of T1 and T3. This shows that for electrode AF3 and the band alpha, there is a U-relation between the RP across tasks T1, T2 and T3.

    \item \textbf{Theta band}: Fig. \ref{fig:graph_distribution_alpha_theta}(b) shows alpha RP across all electrodes and tasks. Considering electrode AF3 and theta band, the mean theta RP is highest for T3, and T2 has the lowest. For T1, the mean theta RP is between that of T3 and T2. This shows that for electrode AF4 and the band theta, there is a U-shaped relation between the RP across tasks T1, T2 and T3. 

    Consider electrode location AF4 and theta band, the mean theta RP is highest for T3,  and T2 has the lowest. For T1, the mean theta RP is between that of T2 and T3. This shows that for electrode AF4 and the band theta, there is a U-shaped relation between the RP across tasks T1, T2 and T3. 
\end{itemize}

    \item \textbf{Electrode-wise analysis}: For electrode location AF3 and the alpha band, the mean RP for task T1 is 21.29\% with SD = 8.65. Similarly, for electrode location AF4 and the alpha band, the mean RP for task T1 is 19.75\% with SD = 9.034. Their difference in the RP 1.54\%. Similarly, differences in means of RPs between AF3 and electrode AF4 locations are shown in Table \ref{tab:AF3-AF4}. Alpha and theta RP distributions are shown in Fig. \ref{fig:graph_distribution_alpha_theta}(a) and \ref{fig:graph_distribution_alpha_theta}(b), respectively. The differences between the means are relatively low, although the significance of the difference is statistically explored in later sections as part of ``within-subjects".

\end{enumerate}

For a chosen electrode and a chosen band, say alpha, the mean RP decreases from task T1 to T2 and then increases from task T2 to T3. In other words, it has a U-shaped relation from tasks T1, T2 and T3. Considering the mean theta RP, there is an increase in its value from task T1 to T2 and a decrease from task T2 to T3. In other words, it has a U-shaped relation from tasks T1, T2 and T3. Clearly, there is a similar pattern of relation between alpha and theta RP observed in the order of T1, T2 and T3. But the U-relation is more prominent in alpha compared to theta. This shows that there is more desynchronization of alpha and less synchronisation of theta from T1 to T3. 

To keep track of tasks, electrodes, and RPs is difficult, so new parameters have been introduced to reduce the dimensionality for easy understanding of the cognitive response of the participants. For this simplification, using the RPs of alpha and theta, inter-BRPD is estimated as mentioned in steps 9 of Table \ref{tab:parameters_estimation} respectively. The inter-BRPD, especially its mean, is an indicative parameter of mental effort.

\subsubsection{Inter-BRPD Analysis}
Inter-BRPD gives an indication of change synchronisation of theta and desynchronisation of alpha. The subtrahends as shown in table \ref{tab:parameters_estimation}, cannot be RPs from the same band, i.e. they need to be of different bands. In the current work, alpha and theta's RPs are the subtrahends. Inter-BRPD is based on the assumption that for an increase in mental effort, there is synchronisation of theta and resynchronisation of alpha simultaneously. In other words, mental effort is negatively correlated to alpha power (or RP) and positively correlated to theta power (or RP). So, ideally, an increase in mental effort is characterised as shown in Table \ref{tab:mental_effort_cases}.

\small
\begin{table}[t]
\centering
\caption{Correlation of change in Mental Effort and inter-BRPD}
\label{tab:mental_effort_cases}
\scriptsize
\begin{tabular}{@{}cllll@{}}
\toprule
\textbf{Case} & \textbf{Mental Effort} & \textbf{Alpha RP} & \textbf{Theta RP} & \textbf{Inter-BRPD} \\ 
\midrule
1 & \multirow{3}{*}{Increase} & Decrease & Increase & \multirow{3}{*}{Decrease}\\
2 &  & No Change & Increase \\
3 &  & Decrease & No Change \\
\midrule
4 & \multirow{3}{*}{Decrease} & Increase & Decrease & \multirow{3}{*}{Increase} \\
5 &  & No Change & Decrease \\
6 &  & Increase & No Change \\
\bottomrule
\end{tabular}
\end{table}

\normalsize
Inter-BRPD of a participant or a task has to be used in comparison with other values of inter-BRPD of another task or participant. For example, if there are two tasks, T$_i$ and T$_j$, with inter-BRPD values as I$_i$ and I$_j$, respectively. Then, the question follows which task has higher mental effort or whether there is an increase or decrease in mental effort between the tasks based on the change in inter-BRPD values. For this, based on the assumptions followed in Table \ref{tab:mental_effort_cases}, it can be concluded that if  I$_i$ $<$ I$_j$, then the mental effort is higher T$_i$. The main reason for this is that inter-BRPD has alpha RP as minuend and theta RP as subtrahend. As alpha RP and theta RP are negatively correlated, an increase in alpha RP is characterised by a decrease in RP, which tends to increase the inter-BRPD value. On the other hand, a decrease in alpha RP is characterised by an increase in theta RP, which tends to decrease the inter-BRPD value. Thus, inter-BRPD can be used to measure mental effort.

\begin{figure}[t]
    \centering
    \includegraphics[width=0.7\linewidth]{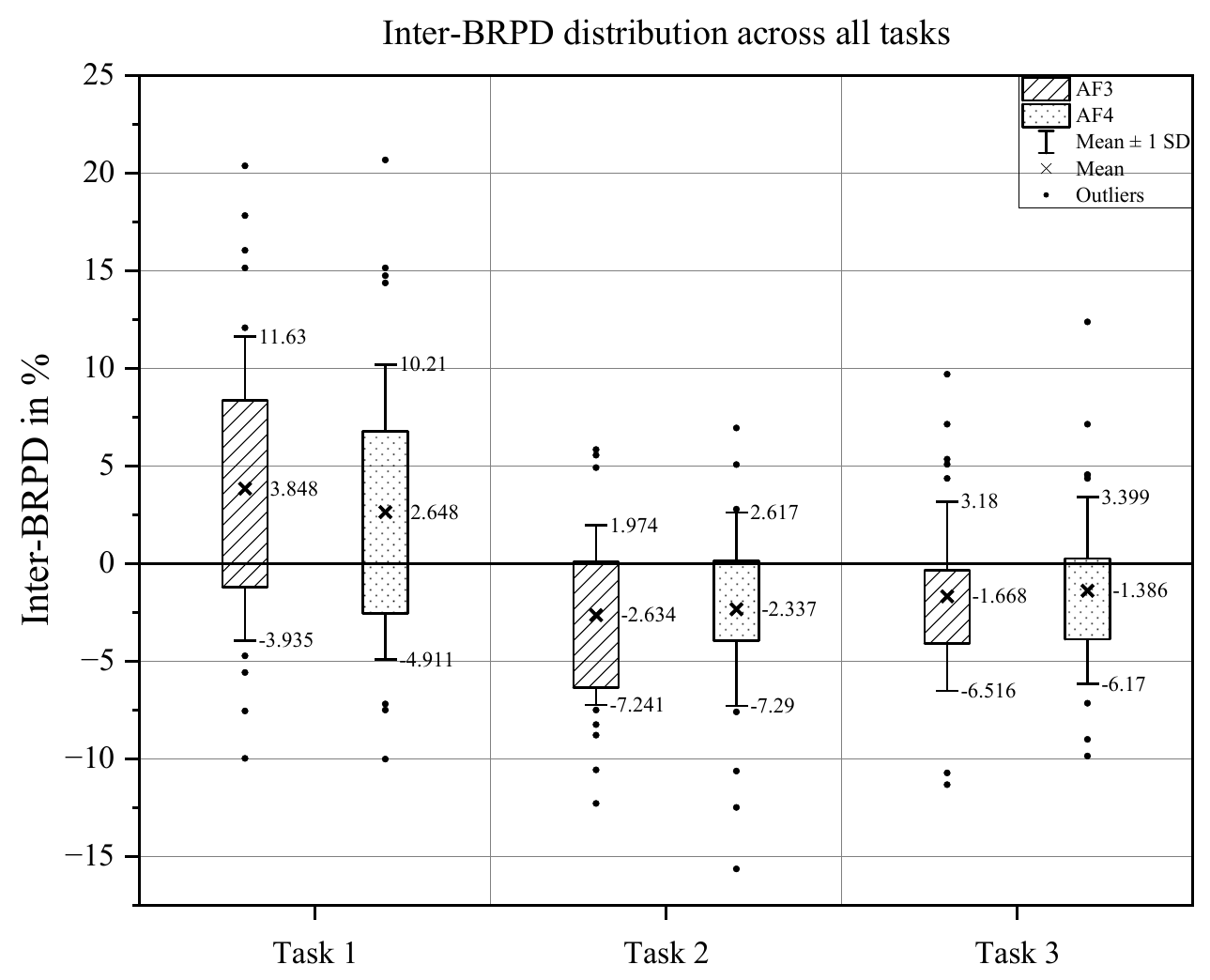}
    \caption{Comparison of inter-BRPD across all tasks.}
    \label{fig:Graph_CPI}
\end{figure}

\begin{table}[t]
\centering
\caption{Descriptive Statistics for inter-BRPD at AF3 and AF4 across Tasks}
\scriptsize
\begin{tabular}{llrrrrrrr}
\toprule
\textbf{Electrode} & \textbf{Task} & \textbf{Mean} & \textbf{CI Low} & \textbf{CI High} & \textbf{Median} & \textbf{SE} & \textbf{SD} & \textbf{Min / Max} \\
\midrule
\multirow{3}{*}{AF3} 
& Task 1 & 3.85 & 0.77 & 6.93 & 2.79 & 1.50 & 7.78 & -9.98 / 20.39 \\
& Task 2 & -2.63 & -4.46 & -0.81 & -2.13 & 0.89 & 4.61 & -12.29 / 5.84 \\
& Task 3 & -1.67 & -3.59 & 0.25 & -2.01 & 0.93 & 4.85 & -11.33 / 9.70 \\
\midrule
\multirow{3}{*}{AF4} 
& Task 1 & 2.65 & -0.34 & 5.64 & 1.76 & 1.45 & 7.56 & -10.01 / 20.69 \\
& Task 2 & -2.34 & -4.30 & -0.38 & -2.17 & 0.95 & 4.95 & -15.65 / 6.95 \\
& Task 3 & -1.39 & -3.28 & 0.51 & -1.96 & 0.92 & 4.78 & -9.86 / 12.38 \\
\bottomrule
\end{tabular}
\label{tab:desc_stats_inter-BRPD}
\end{table}

The Fig. \ref{fig:Graph_CPI} shows the inter-BRPD distribution between alpha and theta for each task and for each electrode location. Table \ref{tab:desc_stats_inter-BRPD} gives descriptive statistics for inter-BRPD for each task and electrode location. For each task, there are two box plots indicating the distribution of inter-BRPD for the two electrodes. The hatched and sparsely dotted box plots refer to electrode locations AF3 and AF4, respectively. The mean of the inter-BRPD for each electrode of each task is indicated by the \textbf{$\times$} mark in the box plots. For electrode AF3, the mean inter-BRPD is positive for T1 and negative for T2 and T3. Moreover, the mean inter-BRPD is lowest in T2 compared to that in T1 and T2.  

Similarly, for electrode location AF4, the mean inter-BRPD is positive for T1 and negative for T2  and T3. Moreover, the mean inter-BRPD is lowest in task T2 compared to that in tasks T1 and T2. On comparing inter-BRPD values between tasks T1 and T2, T2 has a lower inter-BRPD value. This is characterised by the minimal change (almost constant) in theta RP, and there is a drastic decrease in alpha RP. As per table \ref{tab:mental_effort_cases}, there is a mental effort increase from T1 to T2. Also, on comparing inter-BRPD values between tasks T2 and T3, T2 has a lower inter-BRPD value. This is characterised by minimal change (almost constant) in theta RP and a drastic decrease in alpha RP. Overall, one can say that the highest mental workload was in task T2 due highest mean inter-BRPD. The next highest is in task T3 due to the second-highest mean of inter-BRPD. As the mean inter-BRPD is highest in task T1, it shows that task T1 has the lowest mental workload. Here also, the mental workload shows a U-shaped pattern from task T1 to T3. 

\section{Results}
The relative band powers obtained are investigated to determine whether a pattern exists between tasks and electrode locations. For this, at first the data is tested for normality and descriptive statistics are generated. Later, an appropriate statistical test is used to investigate the pattern between tasks and electrode locations. The results section is divided into two parts i.e., one is the statistical validation and correlation with other cognitive indices from literature.

\subsection{Statistical validation}
\label{normalitysection}
Even though the mean of inter-BRPD is evident from the graph, it is necessary to validate the results with statistics.  Here, there are three independent variables (IV) i.e. tasks T1, T2 and T3. These IVs are given as ordinal input data to the participants. There are two dependent variables (DV) are the inter-BRPD for electrode location AF3 and AF4. The DVs are ratio scale data. To test the significance of the results, the general linear model is used in IBM SPSS. The level of significance and confidence interval are 0.05 and 95\% respectively. To validate the significance of the results, Hypothesis 1 mentioned in section \ref{sec:methodology} needs to be tested. 

Before going into the details of statistical analysis, the inter-BRPD data should first be tested for normality. Inter-BRPD data are tested for normality using the Shapiro-Wilk test using IBM SPSS. Table \ref{tab:normality_interBRPD} gives normality test results of inter-BRPD data. Shapiro-Wilk test is mostly suited for small sample sizes as mentioned in \cite{razali2011power}. The p-value $>$ 0 for all data which shows that the null hypothesis is rejected. Thus, all the data generated after the finding inter-BRPD between alpha and theta are normally distributed. Then, the data is tested for its statistical significance for electrode-wise (within-subjects effect) and task-wise (between-subject effect). This implies that statistical tests for significance can be applied using general linear model in IBM SPSS. 

\begin{table}[t]
\centering
\caption{Tests of Normality for inter-BRPD data}
\label{tab:normality_interBRPD}
\scriptsize
\begin{tabular}{llcccccc}
\toprule
\textbf{Task} & & \multicolumn{3}{c}{\textbf{Kolmogorov-Smirnov\textsuperscript{a}}} & \multicolumn{3}{c}{\textbf{Shapiro-Wilk}} \\
\cmidrule(lr){3-5} \cmidrule(lr){6-8}
& & \textbf{Statistic} & \textbf{df} & \textbf{Sig.} & \textbf{Statistic} & \textbf{df} & \textbf{Sig.} \\
\midrule
\multirow{3}{*}{AF3} 
& Task 1 & .124 & 27 & .200\textsuperscript{*} & .971 & 27 & .624 \\
& Task 2 & .121 & 27 & .200\textsuperscript{*} & .972 & 27 & .663 \\
& Task 3 & .170 & 27 & .044 & .947 & 27 & .186 \\
\midrule
\multirow{3}{*}{AF4} 
& Task 1 & .102 & 27 & .200\textsuperscript{*} & .965 & 27 & .466 \\
& Task 2 & .154 & 27 & .098 & .934 & 27 & .084 \\
& Task 3 & .167 & 27 & .053 & .929 & 27 & .067 \\
\bottomrule
\end{tabular}
\vspace{0.5em}
\begin{flushleft}
\textsuperscript{*} This is a lower bound of the true significance.\\
\textsuperscript{a} Lilliefors Significance Correction
\end{flushleft}
\end{table}

\begin{table}[t]
\centering
\caption{Tests of Within-Subjects Contrasts}
\label{tab:within_subjects_contrasts}
\scriptsize
\begin{tabular}{llcccccc}
\toprule
\textbf{Source} & \textbf{Electrode\_location} & \textbf{Type III Sum of Squares} & \textbf{df} & \textbf{Mean Square} & \textbf{F} & \textbf{Sig.} \\
\midrule
Electrode\_location & Linear & 1.731 & 1 & 1.731 & .583 & .447 \\
Electrode\_location * Task & Linear & 19.958 & 2 & 9.979 & 3.361 & .040 \\
Error(Electrode\_location) & Linear & 231.587 & 78 & 2.969 & -- & -- \\
\bottomrule
\end{tabular}
\end{table}

\begin{table}[t]
\centering
\caption{Tests of Between-Subjects Effects}
\label{tab:between_subjects_effects}
\scriptsize
\begin{tabular}{lccccc}
\toprule
\textbf{Source} & \textbf{Type III Sum of Squares} & \textbf{df} & \textbf{Mean Square} & \textbf{F} & \textbf{Sig.} \\
\midrule
Intercept & 10.508 & 1 & 10.508 & 0.157 & 0.693 \\
Task      & 1018.559 & 2 & 509.280 & 7.603 & $<.001$ \\
Error     & 5224.927 & 78 & 66.986 & -- & -- \\
\bottomrule
\end{tabular}
\end{table}

\begin{table}[ht]
\centering
\caption{Post Hoc Tests – Bonferroni Multiple Comparisons}
\label{tab:multiple_comparisons}
\scriptsize
\begin{tabular}{llcccccc}
\toprule
\textbf{(I) Task} & \textbf{(J) Task} & \textbf{Mean Difference (I-J)} & \textbf{Std. Error} & \textbf{Sig.} & \textbf{95\% CI Lower Bound} & \textbf{Upper Bound} \\
\midrule
\multirow{2}{*}{Task 1} 
    & Task 2 & \textbf{5.733*} & 1.575 & \textbf{0.001} & 1.879 & 9.586 \\
    & Task 3 & \textbf{4.774*} & 1.575 & \textbf{0.010} & 0.920 & 8.628 \\
\midrule
\multirow{2}{*}{Task 2} 
    & Task 1 & \textbf{-5.733*} & 1.575 & \textbf{0.001} & -9.586 & -1.879 \\
    & Task 3 & -0.958 & 1.575 & 1.000 & -4.812 & 2.895 \\
\midrule
\multirow{2}{*}{Task 3} 
    & Task 1 & \textbf{-4.774*} & 1.575 & \textbf{0.010} & -8.628 & -0.920 \\
    & Task 2 & 0.958 & 1.575 & 1.000 & -2.895 & 4.812 \\
\bottomrule
\end{tabular}
\vspace{1ex}
\small
\begin{minipage}{0.9\linewidth}
\textit{Note:} Based on observed means. \\
The error term is Mean Square(Error) = 33.493. \\
* The mean difference is significant at the .05 level.
\end{minipage}
\end{table}

\normalsize

\begin{itemize}
    \item \textbf{Betweeen subjects effects}: Table \ref{tab:between_subjects_effects} shows between-subject effects. This shows there is a significant difference between tasks (F(2,78) = 7.603, p $<$ .001). This is also evident from the assumptions that there should be a difference between the tasks. Post-hoc analysis is performed using a Bonferroni-corrected test, which gives details about the significant differences between tasks T1, T2 and T3 pairwise. Table \ref{tab:multiple_comparisons} shows the pairwise comparison differences. From this, it can be concluded that task T1 is significantly different (p $<$0.05) from tasks T2 and T3. Also, there is no significant difference (p$>$0) between task T2 and T3. This is true from the design experiment also. Task T1 is a simple points joining task and whereas tasks T2 and T3 are cognitively involved. Also, there is not much of a significant difference between tasks T2 and T3. This answers the research question RQ1.
    
    \item \textbf{Within subjects effects}: Table \ref{tab:within_subjects_contrasts} shows within-subject effects. This implies that the electrode location itself had no significant effect (p = 0.447 $>$ 0.05), which shows that the differences between electrode locations AF3 and AF4 were minimal or absent. Hence, during the tasks, instead of two,  one electrode can be used for measuring the mental effort during a design activity. This answers the research question RQ2 as electrode locations AF3 and AF4 data are related to the left and right brain hemispheres respectively, in terms of cognition. On the other hand, the interaction between electrode and task says that the difference between electrode locations varied significantly (F(2,78) = 3.361, p = .04 $<$ 0.05) based on the task. This implies that there is an effect of the task on the electrode. 
\end{itemize}

\subsection{Importance of inter-BRPD}
This section gives an overview of the potential of the parameter inter-BRPD as an estimate of mental effort. There are other parameters, as mentioned in the literature, such as Cognitive load index, excitement index, relaxation index, engagement index and mental fatigue. Their formulae are mentioned in \cite{blanco2024real}. Alpha and theta RPs are also one of the parameters to measure mental effort or cognitive load. In this work, a correlation matrix is estimated to find out which one of the parameters gives the closest and practical relationship with alpha and theta power distribution. Here, the data collected during the experiment is used to find the correlation matrices. As mentioned earlier in the methods and materials section, there were 3 tasks, and in each task the response is estimated at two electrode locations, AF3 and AF4. For each parameter, the distribution of the indices across the electrode locations and tasks is shown in Fig. \ref{fig:Indices_CL_distribution} - \ref{fig:Indices_Relaxation_distribution}.

In this work, the correlation is found between the 8 variables, i.e., alpha RP, theta RP, relaxation index, excitement index, mental fatigue index, cognitive load index and inter-BRPD. As there are two electrode locations and 3 tasks, 6 correlation matrices are obtained using IBM SPSS. Using MATLAB, all six correlation matrices are combined using Fischer's z-transform \cite{fiser_zTranform}. The MATLAB code and the correlation matrices are shown in Appendix B. 

\begin{figure}
    \centering
    \includegraphics[width=0.5\linewidth]{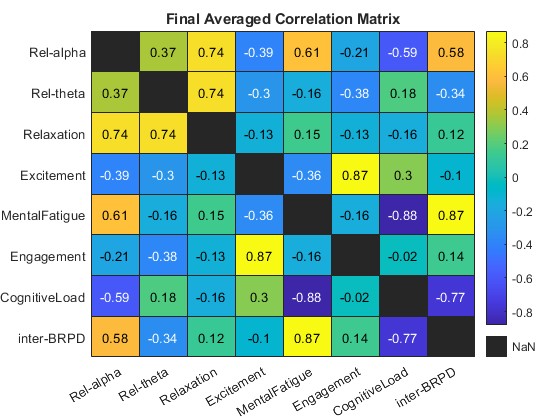}
    \caption{Correlation matrix heat map}
    \label{fig:heatmap}
\end{figure}

The output of the MATLAB code is the overall correlation matrix which is shown in the form of heatmap in Fig. \ref{fig:heatmap}. The values of the matrices are the Pearson correlation values. They range from -0.8 to 0.8. The most positively correlated value is 0.87 between the mental fatigue index and inter-BRPD. The most negatively correlated value is -0.88 between the mental fatigue index and the cognitive load index. But the comparison should be done with relative alpha and theta RPs because they are the reference for us. 

For alpha RP, the highest correlated value is 0.74 with the relaxation index. The least correlated value is -0.59 with the cognitive load index. Similarly, for theta RP, the highest positively correlated value is 0.74 with the relaxation index, and the least negatively correlated value is -0.38 with the engagement index. But there is no consistency with the sign of the parameters of relaxation index, excitement and engagement, so these can be discounted from considerations. Consistency can be checked from Table \ref{tab:mental_effort_cases}. This pattern is not observed in relaxation, excitement and engagement indices. 

The remaining parameters are mental fatigue, cognitive load index and inter-BRPD. On observation, the mental fatigue index and the cognitive load index are reciprocal to each other. Out of these, the mental fatigue load index gives more promising correlation values because of the sign. So, on comparing the mental fatigue index and inter-BRPD, the latter shows more correlation values. This shows that the inte-BRPD is more sophisticated than all other cognitive parameters.

\section{Discussion}
The current work gives a novel, comprehensive understanding of measuring mental effort in motion exploration and concept generation tasks using a new EEG-based metric. For this, three tasks, T1, T2 (MET) and T3 (CGT), were given to participants, and their brain responses were recorded using EEG. In this study, T1 was the control task, and T2 and T3 were the focus tasks. The collected data in the form of raw EEG signals are cleaned, processed and analysed. Finally, the data was made ready to find out the RPs of each band. For both the electrode locations AF3 and AF4, there was a U-shaped relation between the RP of alpha or theta, but the U-shape was more prominent in alpha \cite{klimesch1999eeg,jensen2002frontal,sammer2007relationship}. This was contrary to the belief that the concept generation task had the highest mental effort/workload as mentioned in the literature \cite{nikulin2019nasa} in the conceptual stage of design process. 

For each task, RPs were calculated for alpha and theta at AF3 and AF4 electrode locations. This implies that four RP values are estimated per task for each participant, resulting in a total of 12 RP values across the three tasks. Comparing RPs of similar bands across electrode locations and tasks adds to the complexity due to high dimensionality. To address this, a new EEG-based metric called inter-BRPD is derived, defined as the difference in RPs of alpha and theta bands for a given task and an electrode location. By using this metric, there is a 50\% dimensionality reduction of the data, i.e., the number of values to be compared across tasks and electrode locations decreases from 12 to 6 (check the number of rows in Table \ref{tab:desc_stats_RP} and \ref{tab:desc_stats_inter-BRPD} for each participant.

The mean of the inter-BRPD gives an understanding of the mental workload of the sample of participants when they have undergone tasks T1, T2 and T3. As observed, this mean was lowest in task 2, which shows that there was the highest desynchronisation of alpha and minimum synchronisation of theta. This shows that the mental effort during MET, even though it was statistically similar to CGT, is greater than that of CGT. Perhaps, a higher sample size would show a statistically significant difference in the nature of the tasks. This implies that the amount of mental workload in CGT is comparable to that in MET. Hence, there should be some efforts in finding out ways to identify hotspots among the design actions to reduce the mental effort during the conceptual design phase, which is considered in the future scope. 

For a given task, statistically, there is no difference between the inter-BRPD values at AF3 and AF4; hence, one electrode would suffice instead of two electrodes for gathering EEG data for such a study. Perhaps a larger sample size would reveal a significant difference. But if the means of inter-BRPDs are compared, the mental effort observed during MET is more than that of CGT. This is because MET is more logical, analytical, and reality-based. In the case of CGT, the designers were asked to explore or generate concepts based on the functionality, leveraging creative freedom. However, since the mental effort, as indicated by the observed inter-BRPD, showed a decrease in CGT, it can be inferred that the creative and imaginary aspects of CGT may have led to cognitive relaxation. This is in contrast to the behaviour observed during the MET, where the participants experienced a higher mental effort due to the constrained nature of the task. Studying motion exploration from a creative, non-physics-based, or abstract perspective using inter-BRPD is a promising direction for future research.

Since the new metric was introduced, it needed validation for its usefulness as compared to existing metrics mentioned in \cite{blanco2024real}. 
The correlation analysis performed between inter-BRPD and other cognitive parameters showed that it is more useful in quantifying mental effort. Theoretically, there can be a general expression to combine all band powers to find a unified cognitive parameter to measure cognitive processing in the form of the following equation. 

\begin{eqnarray}
\label{eqn:CognitiveIndexFormula}
    \text{Index} & = k_0 + k_1 \times \text{rel\_alpha\_power} + k_2 \times \text{rel\_theta\_power} +  \\ \notag
    &k_3 \times \text{rel\_delta\_power} +  k_4 \times \text{rel\_beta\_power} \text{ $\forall$ $k_i$ $\in$ $\mathbb{R}$}
\end{eqnarray}

 where $i=0, 1, 2, 3, 4$. $k_1, k_2$, and $k_3$ values can be predicted for the choice of the curve fit of the data. A combination of all such operators can derive a unified possible formula for the cognitive index. For the present study, $k_0=k_3 = k_4 = 0, k_1 = 1$ and $k_2 = -1$ give the difference of the RP of alpha and theta. The value of $ k_i$ can be found out using any optimisation or any machine learning algorithm to mathematically relate the equation to cognitive processing.

One of the limitations of this study is that it takes into account EEG variations across the frontal lobe only, without considering the effects of the tasks on other lobes. This introduces a future scope of looking at EEG variations across different lobes with respect to Inter-BRPD for varying complexity of tasks. Statistically, there is no difference between T2 and T3 in terms of mental effort. However, for  a larger sample size, tighter confidence interval and level of significance differences in T2 and T3 could be observed. Manual artefact rejection of EEG signal is tedious. An automated ML based algorithm for artefact detection may be reliable and efficient for data cleaning process.

\section{Conclusion}
The current study gives a novel EEG-derived metric, called inter-BRPD, to estimate the mental effort experienced by the designers when they are sketching concepts of articulated products. The designers were given a convergent thinking activity and a divergent thinking activity called MET and CGT, respectively. As per this study, the mental effort is high in MET compared to CGT because the former is a highly constrained activity. Hence, it was cognitively loaded. The inter-BRPD is negatively correlated to the frontal-alpha band and positively correlated to the frontal-theta band. The efficacy of the inter-BRPD in relation to other literature-based cognitive parameters is analysed statistically in the form of a consolidated correlation matrix. The validation of the new EEG-metric derived in this study has the potential to be extended to other DFX \cite{Andrew_dfx_2013} tasks to get an overall view of mental effort in the conceptual design phase to find the hotspots.  

\section*{Acknowledgements}
Partially supported by the Department of Science and Technology (DST), Science and Engineering Research Board (SERB) (File no. CRG/2020/005334).

\section*{Declarations}
\begin{itemize}
\item \textbf{Funding}: Partially supported by the Department of Science and Technology (DST),  Science and Engineering Research Board (SERB) (File no. CRG/2020/005334).
\item \textbf{Conflict of interest/Competing interests:} The authors have no competing interests to declare that are relevant to the content of this article.
\item \textbf{Ethics approval and consent to participate:} As per the Indian Council of Medical Research, National Ethical Guidelines for Biomedical \& Health Research Involving Human Participants, New Delhi: ICMR; 2017. Section 4, Page 35, Box 4.4(b) ethical approval has been taken. All participants provided informed consent prior to their participation, and their data were anonymised to maintain confidentiality.
\item \textbf{Consent for publication}: All authors have given consent for publication.
\item \textbf{Data availability}: Not applicable
\item \textbf{Materials availability}: Not applicable
\item \textbf{Code availability}: Not applicable
\item \textbf{Use of AI}: ChatGPT 4.0 was used to produce LaTeX table environments and improve the English language in the manuscript. 
\item \textbf{Author contribution}

\begin{enumerate}
    \item \textbf{G. Kalyan Ramana}: Idea inception about mental effort during design process. Contributed to experimental design, data collection, and inter-BRPD derivation. Responsible for data analysis and statistical validation. Contributed to the creation of all figures except Figures 3–5. Participated in writing and finalizing the Introduction, Methodology, Discussion, and Conclusion.
    \item \textbf{Sumit Yempalle}: Idea inception about measuring mental effort using EEG. Contributed to experimental design, data collection, and inter-BRPD derivation. Responsible for EEG data cleaning, preprocessing, and feature extraction. Created Figures 3–5. Participated in writing and finalizing the Introduction, Methodology, Discussion, and Conclusion.
    \item \textbf{Dr. Prasad S. Onkar}: Overall coordination of the research work. Provided technical guidance throughout the study and reviewed the manuscript for grammatical and scientific accuracy.
\end{enumerate}
\end{itemize}

\newpage
\begin{appendices}
\section{Distribution of additional cognitive parameters}




\begin{figure}[h]
    \centering
    \begin{subfigure}[b]{0.3\linewidth}
        \centering
        \includegraphics[width=\linewidth]{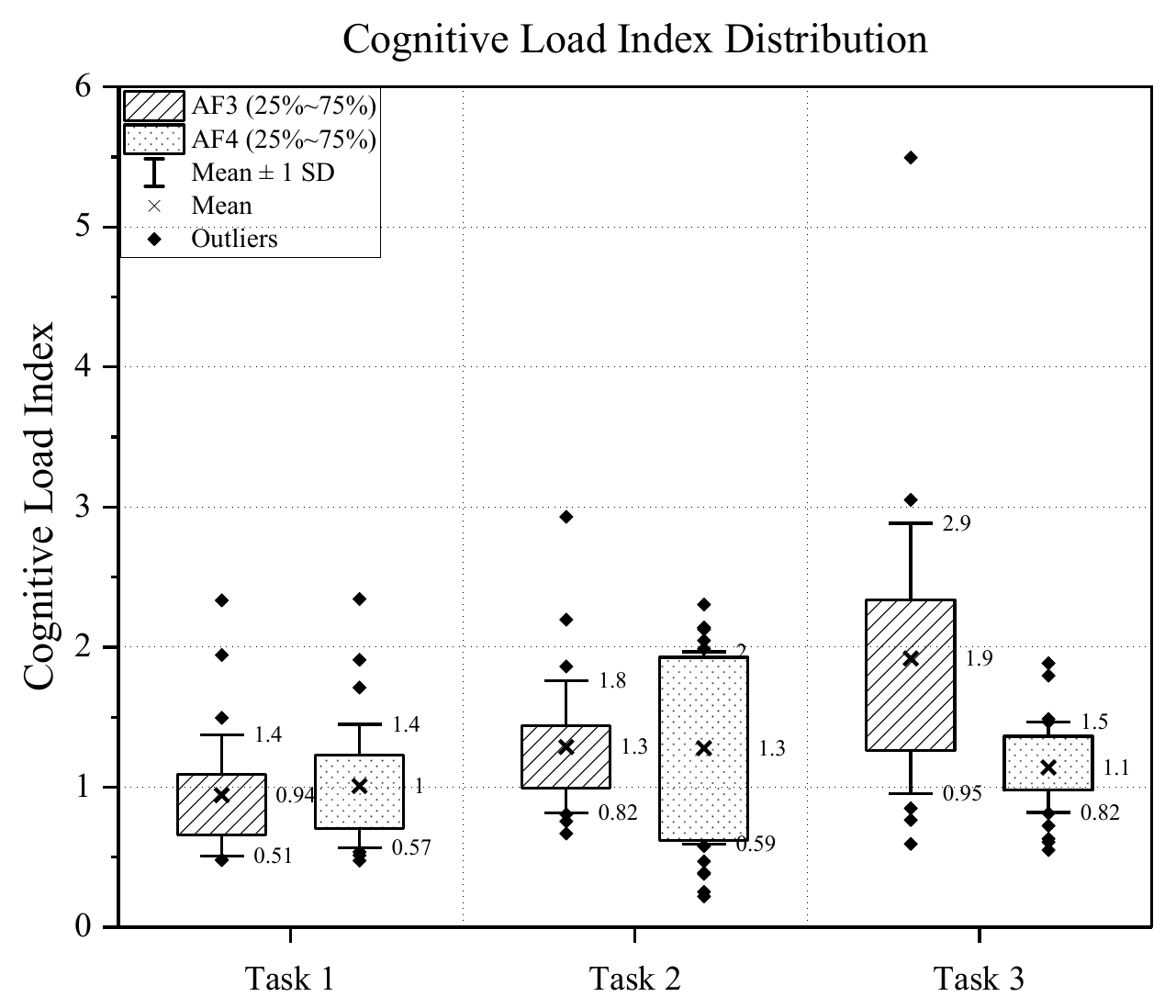}
        \caption{Cognitive Load Distribution}
        \label{fig:Indices_CL_distribution}
    \end{subfigure}
    \quad
    \begin{subfigure}[b]{0.3\linewidth}
        \centering
        \includegraphics[width=\linewidth]{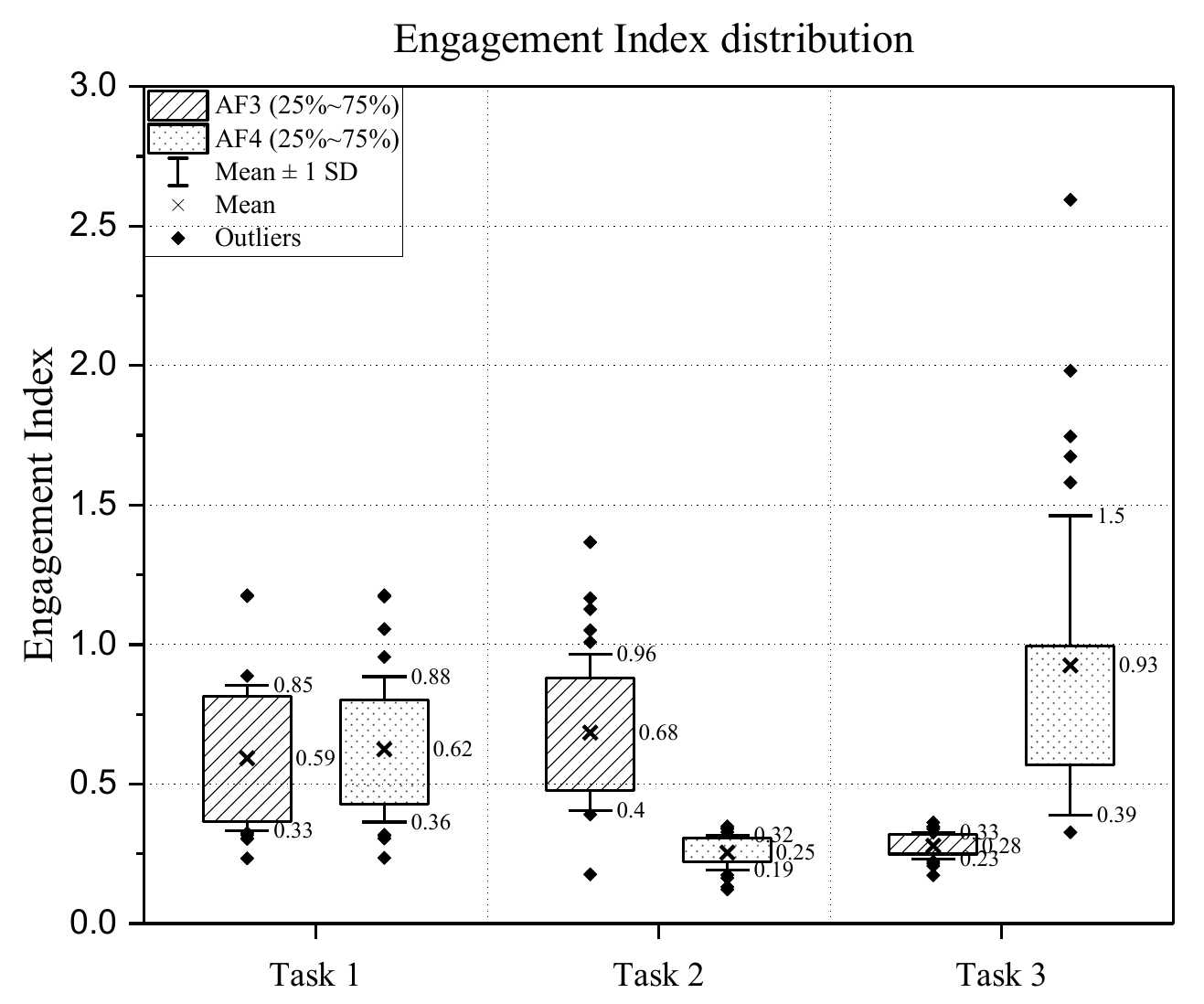}
        \caption{Engagement Index Distribution}
        \label{fig:Indices_Engagement_distribution}
    \end{subfigure}

    \vspace{0.4cm}

    \begin{subfigure}[b]{0.3\linewidth}
        \centering
        \includegraphics[width=\linewidth]{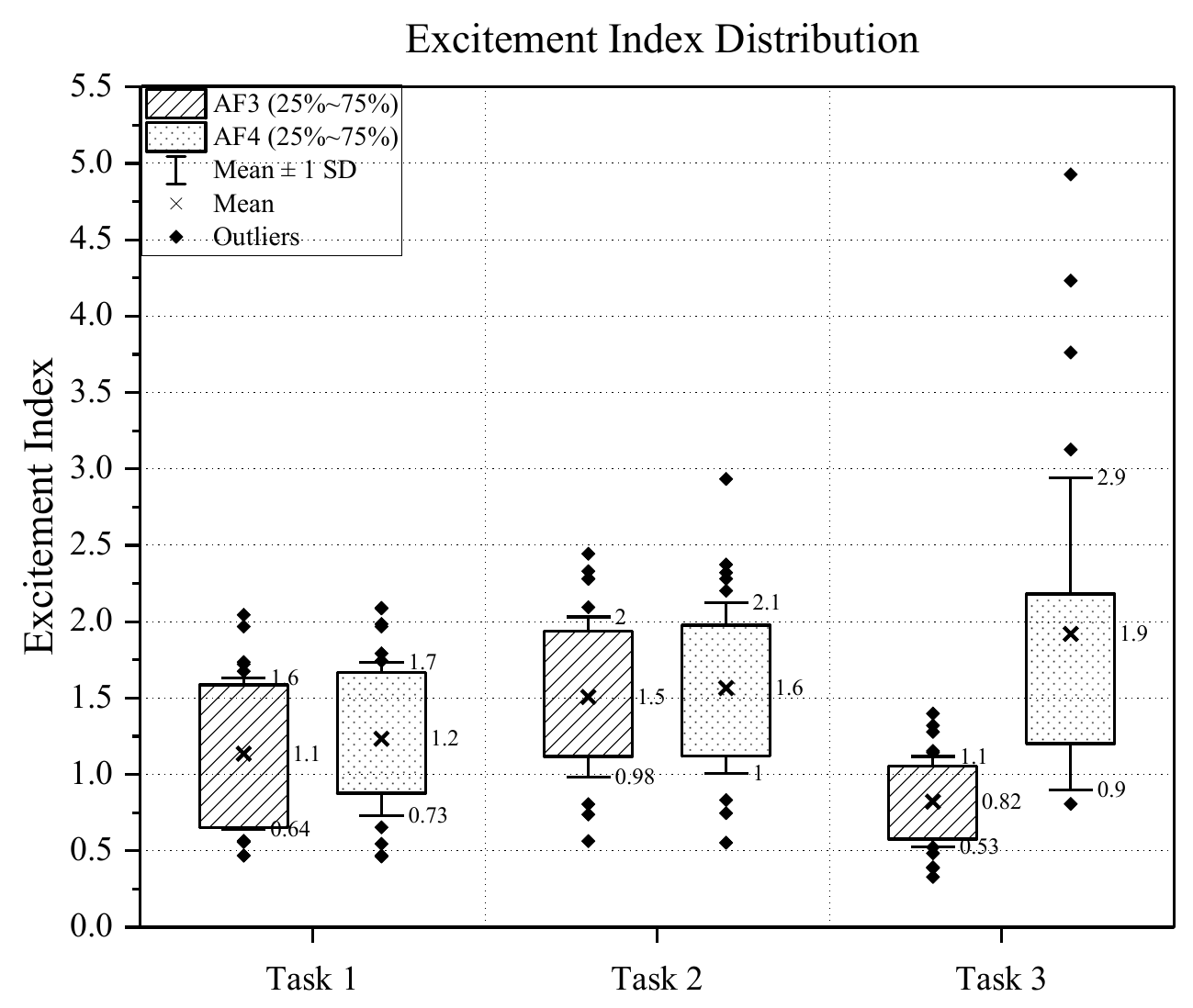}
        \caption{Excitement Index Distribution}
        \label{fig:Indices_Excitement_distribution}
    \end{subfigure}
    \quad
    \begin{subfigure}[b]{0.3\linewidth}
        \centering
        \includegraphics[width=\linewidth]{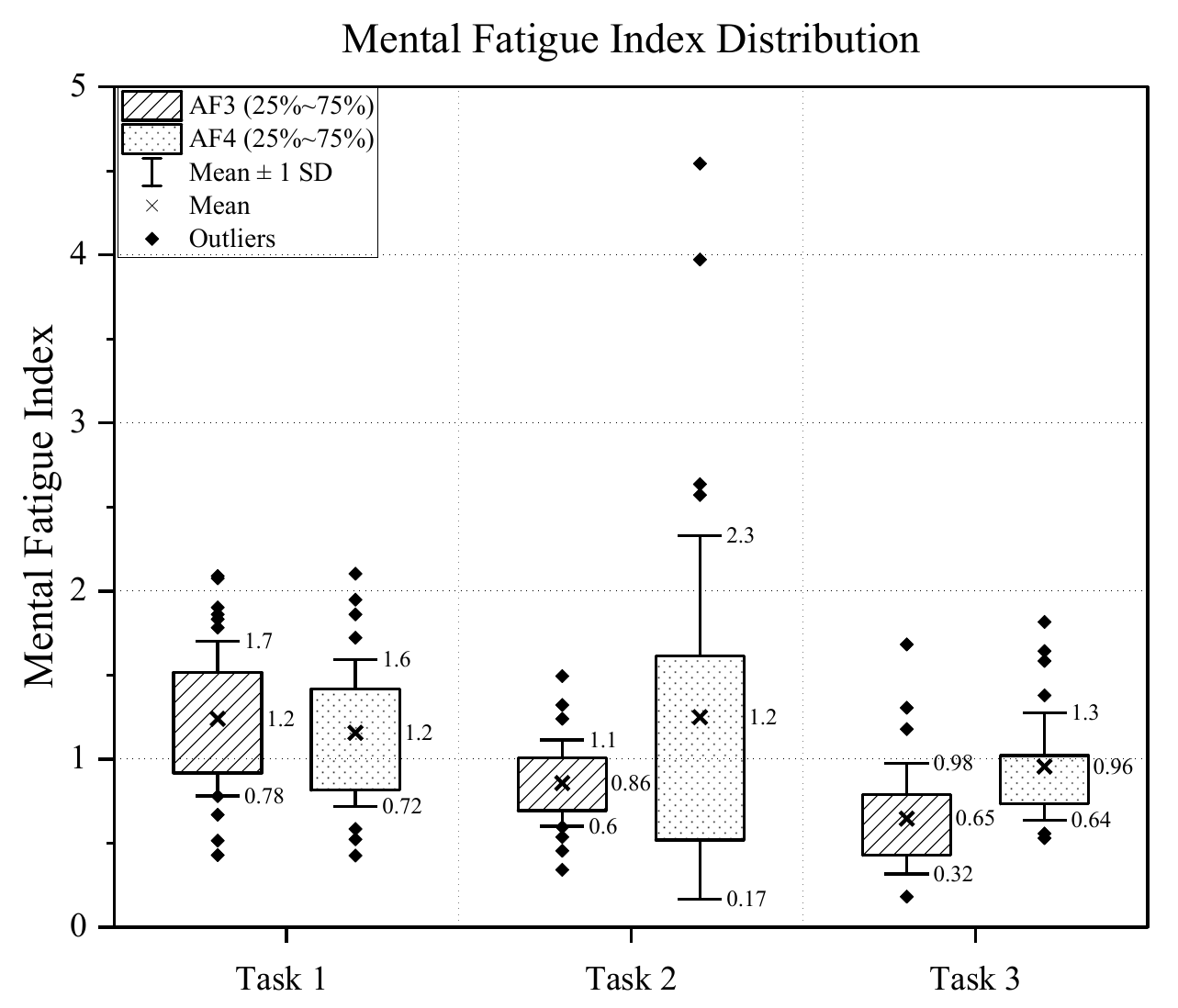}
        \caption{Mental Fatigue Distrbution}
        \label{fig:Indices_MentalFatigue_distribution}
    \end{subfigure}

    \vspace{0.4cm}

    \begin{subfigure}[b]{0.3\linewidth}
        \centering
        \includegraphics[width=\linewidth]{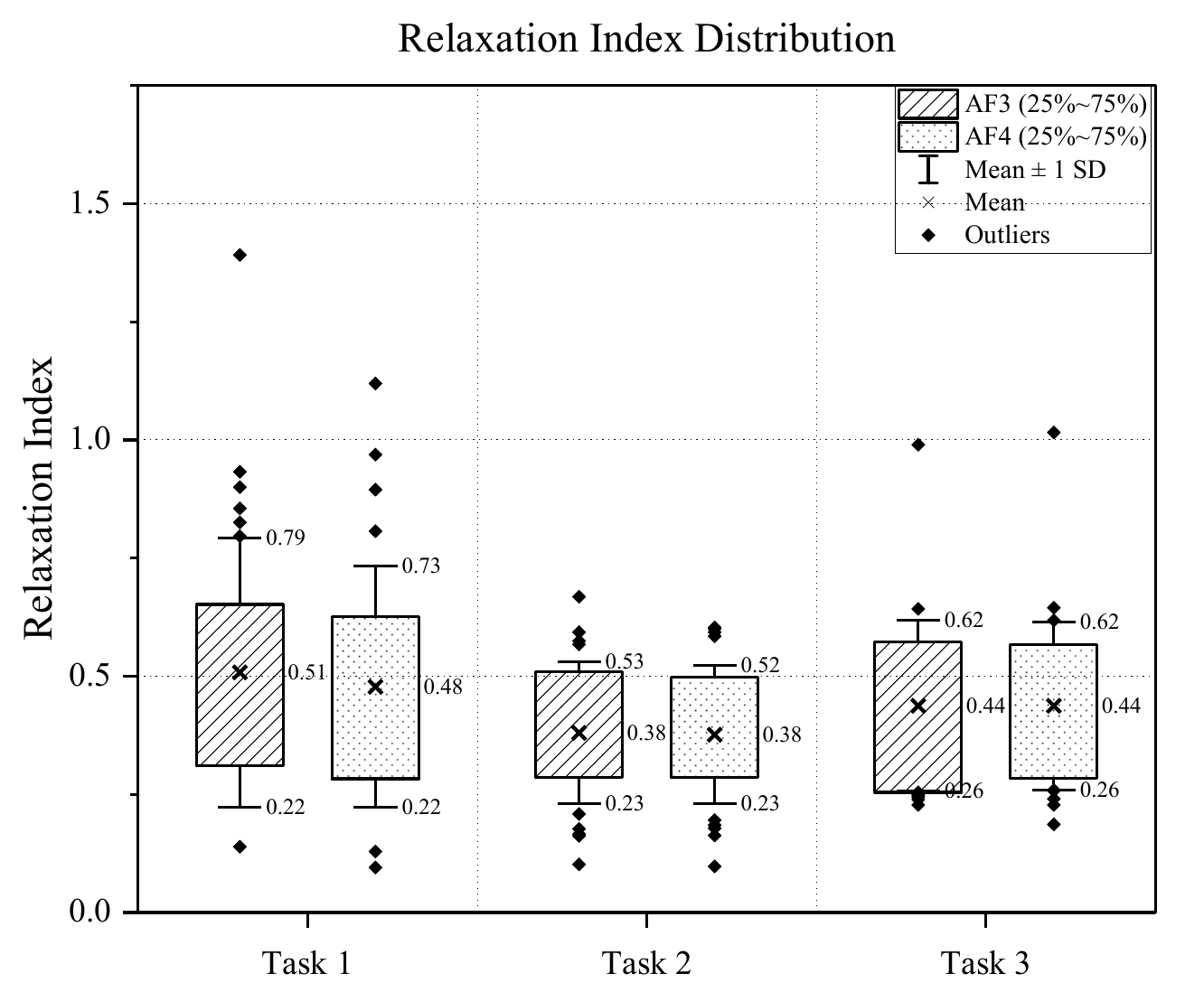}
        \caption{Relaxation Index Distribution}
        \label{fig:Indices_Relaxation_distribution}
    \end{subfigure}

    \caption{Distribution of Cognitive State Indices}
    \label{fig:Indices_All_distribution}
\end{figure}
\end{appendices}
\clearpage
\bibliography{Manuscript_R01}
\end{document}